\documentclass[11pt]{article}

\usepackage{xr}
\usepackage{epsfig}
\usepackage{latexsym}
\usepackage{enumerate}
\usepackage{url}
\usepackage{float}
\usepackage[hypertexnames=false,hyperfootnotes=false]{hyperref}
\usepackage{texnansi}
\usepackage{color}
\usepackage{tikz}
\usepackage[margin=10pt,font=small,labelfont=bf]{caption}
\usepackage[subrefformat=parens,labelformat=parens]{subcaption}
\usepackage{afterpage}
\usepackage{enumitem}
\usepackage[boxed]{algorithm}
\usepackage{algpseudocode}
\usepackage[normalem]{ulem}
\usepackage{lmodern}
\usepackage{booktabs}
\usepackage{sectsty}
\usepackage{ifthen}
\usepackage{amsmath}
\usepackage{amsthm}
\usepackage{amssymb}
\usepackage{amsfonts}
\usepackage{thmtools}
\usepackage{thm-restate}
\usepackage{forest}
\usepackage{mathtools}
\usepackage{xspace}
\usepackage{titling}
\usepackage{listings}
\usepackage{natbib}
\usepackage{xfrac}
\usepackage{comment}
\usepackage{multirow}
\usepackage{bigdelim}
\usepackage{bm}
\usepackage{cleveref}
\usepackage[textsize=tiny]{todonotes}
\newcommand{\field}[1]{\ensuremath{\mathbb{#1}}}
\newcommand{\R}{\ensuremath{\field{R}}} 
\newcommand{\Rp}{\ensuremath{\R_+}} 
\newcommand{\tends}{\ensuremath{\rightarrow}} 
\newcommand{\E}{\ensuremath{\mathsf{E}}} 
\newcommand{\defeq}{\ensuremath{\triangleq}}
\newcommand{\subjectto}{\text{\rm subject to}} 

\newcommand{\Cscr}{\ensuremath{\mathcal C}}

\newcommand{\Fscr}{\ensuremath{\mathcal F}}

\DeclareMathOperator{\diag}{diag}

\DeclareMathOperator{\Var}{Var}

\DeclareMathOperator{\tr}{tr}

\newcommand{\minimize}{\ensuremath{\mathop{\mathrm{minimize}}\limits}}

\declaretheoremstyle[headfont=\sffamily\bfseries,bodyfont=\itshape]{thm-sf}
\declaretheorem[style=thm-sf]{theorem}
\declaretheorem[style=thm-sf]{remark}
\declaretheorem[style=thm-sf]{assumption}
\crefname{assumption}{assumption}{assumptions}

\declaretheorem[style=thm-sf]{example}

\declaretheorem[style=thm-sf]{corollary}
\declaretheorem[style=thm-sf]{lemma}

\declaretheorem[style=thm-sf]{proposition}

\newcommand{\proofnamest}[1]{{\normalfont\sffamily\bfseries #1}}
\renewcommand{\thmcontinues}[1]{\hyperref[#1]{continued}}
\newcommand{\paraheader}[1]{%
  \noindent{\sffamily\bfseries #1}}
\usetikzlibrary{arrows,patterns,plotmarks,pgfplots.groupplots}
\usetikzlibrary{math}
\tikzstyle{every picture} += [>=stealth]
\tikzset{axis/.style={semithick, line join=miter}}
\allsectionsfont{\sffamily}
\makeatletter
\def\@seccntformat#1{\csname the#1\endcsname.\quad}
\makeatother

\floatname{algorithm}{\normalfont\sffamily\bfseries Algorithm}
\floatstyle{ruled}
\newcommand{\emailhref}[1]{\href{mailto:#1}{\tt #1}} 
\provideboolean{fastcompile}
\newcommand{\hidefastcompile}[1]{\ifthenelse{\boolean{fastcompile}}{}{#1}}
\usepackage{pgfplots}
\usepackage{pgfplotstable}
\usetikzlibrary{calc}
\usepackage{mathtools}
\definecolor{orange}{rgb}{0.85,0.33,0.13} 
\definecolor{green}{rgb}{0.13,0.85,0.33}
\definecolor{purple}{rgb}{0.33,0.13,0.85}
\definecolor{lime}{rgb}{0.65,0.85,0.13}
\definecolor{blue}{rgb}{0.13,0.65,0.85}
\pgfplotscreateplotcyclelist{tricolor}{%
  orange,every mark/.append style={fill=orange!80!black},mark=*\\%
  green,every mark/.append style={fill=green!80!black},mark=square*\\%
  purple,every mark/.append style={fill=purple!80!black},mark=otimes*\\%
  black,mark=star\\%
  orange,every mark/.append style={fill=orange!80!black},mark=diamond*\\%
  green,densely dashed,every mark/.append style={solid,fill=green!80!black},mark=*\\%
  purple,densely dashed,every mark/.append style={solid,fill=purple!80!black},mark=square*\\%
  black,densely dashed,every mark/.append style={solid,fill=gray},mark=otimes*\\%
  orange,densely dashed,mark=star,every mark/.append style=solid\\%
  green,densely dashed,every mark/.append style={solid,fill=green!80!black},mark=diamond*\\%
}
\pgfplotsset{colormap={tricolormap}{color=(orange) color=(green) color=(purple)},
  colormap={quadcolormap}{color=(orange) color=(lime) color=(blue) color=(purple)}}
\pgfplotstableset{%
  font=\small,
  every head row/.style={before row=\toprule[1pt], after row=\midrule},
  every last row/.style={after row=\bottomrule[1pt]}}
\pgfplotsset{compat=1.15}

\usepackage{setspace}
\usepackage[margin=1in]{geometry}

\provideboolean{submissionversion}
\setboolean{submissionversion}{true}
\provideboolean{blindversion}

\ifthenelse{\boolean{submissionversion}}{
  \renewcommand{\todo}[2][1]{}
  
  \newcommand{\deledit}[1]{}
}{
  
  \newcommand{\deledit}[1]{{\color{orange} \sout{#1}}}
}

\newcommand{\pnl}{\ensuremath{\mathsf{P\&L}}\xspace}
\newcommand{\LVR}{\ensuremath{\mathsf{LVR}}\xspace}
\newcommand{\LVH}{\ensuremath{\mathsf{LVH}}\xspace}
\newcommand{\LVB}{\ensuremath{\mathsf{LVB}}\xspace}

\newcommand{\Q}{\ensuremath{\mathbb{Q}}\xspace}
\newcommand{\HODL}{\ensuremath{\mathsf{HODL}}\xspace}
\newcommand{\ARB}{\ensuremath{\mathsf{ARB}}\xspace}
\newcommand{\FEE}{\ensuremath{\mathsf{FEE}}\xspace}
\newcommand{\CFMM}{\ensuremath{\mathsf{CFMM}}\xspace}

\ifthenelse{\boolean{blindversion}}{
  \title{\textsf{\textbf{Automated Market Making and Loss-Versus-Rebalancing}}}
  \author{}
  \date{}
}{
  \title{\textsf{\textbf{Automated Market Making and Loss-Versus-Rebalancing}}%
      \thanks{
            The second author thanks Richard Dewey, Craig Newbold, Guillermo Angeris, Tarun
            Chitra, and Alex Evans for helpful conversations on automated market making. We are
            also grateful to Austin Adams, Jun Aoyagi, Eric Budish, Larry Glosten, Gur Huberman, Mingxuan He,
            Thomas Rivera, Xin Wan, and Tianyi Zhang for helpful comments. The second author was
            supported by the Briger Family Digital Finance Lab at Columbia Business School and is
            an advisor to fintech companies including Paradigm, a venture capital firm with
            investments in automated market making protocols. The third author is Head of Research at a16z Crypto, a venture capital firm with investments in automated market making protocols. The first author was supported in part by NSF award CNS-2212745, and in part by an unrestricted gift from Gnosis, Ltd. The third author was supported in part by NSF awards CCF-2006737 and CNS-2212745.
          }
        }
    \author{
      Jason Milionis \\
      Department of Computer Science \\
      Columbia University \\
      \emailhref{jm@cs.columbia.edu} \\
      \and
      Ciamac C. Moallemi \\
      Graduate School of Business \\
      Columbia University \\
      \emailhref{ciamac@gsb.columbia.edu} \\
      \and
      Tim Roughgarden \\
      Department of Computer Science \\
      Columbia University \\
      a16z Crypto \\
      \emailhref{tim.roughgarden@gmail.com} \\
      \and
      Anthony Lee Zhang \\
      Booth School of Business \\
      University of Chicago \\
      \emailhref{anthony.zhang@chicagobooth.edu}
    }
  \date{Initial version: July 31, 2022 \\
    Current version: May 3, 2026
  }
}

\begin{document}
\maketitle
\singlespacing

\begin{abstract}

Automated Market Makers (AMMs) are both liquidity sources and investment vehicles for market participants. This paper analyzes the risks and returns of liquidity provision (LP) investments in AMMs. In a continuous-time model, we show that LP returns decompose into a beta-like component reflecting market risk exposure, and an alpha-like component reflecting microstructural forces: accrued fees minus losses to arbitrageurs. Applying our decomposition to the Uniswap v2 ETH-USDC pool, we find that over 99.991\% of LP return variance is driven by beta exposure to market risk.

\end{abstract}

\doublespacing

\section{Introduction}

In traditional financial markets, retail traders are generally liquidity takers, transacting against quotes set by large, specialized market-making firms. These firms earn profits through bid-ask spreads, which represent transaction costs for retail investors. In decentralized financial markets, \emph{automated market makers (AMMs)} take on many of the functions traditionally performed by market-making firms. Traders can interact with AMMs as liquidity takers, executing trades at posted prices. But unlike in traditional markets, users can also become liquidity providers (LPs) by contributing capital to the AMM\textquoteright s pools. In return, they earn a share of trading fees and bear the associated risks. AMMs thus play a dual role in DeFi: they are both trading venues and investment vehicles.

A growing literature studies AMMs from a microstructural perspective, analyzing how fee-paying noise traders and informed arbitrageurs impact AMM profitability. This paper asks a complementary question: how do automated market makers behave as \emph{investments} for liquidity providers? What are the risks and returns of LP investments in AMMs, and how do they compare to investing directly in the assets held and traded by AMMs?

A core principle in asset pricing is that the returns of complex risky assets can be decomposed into simpler components. A stock's return reflects a combination of factor risk premia and a stock-specific alpha; a corporate bond splits into the risk-free rate, term premia, and a credit risk premium; an option embeds directional and volatility exposures to the underlying stock. These decompositions allow us to think of asset returns as a combination of terms common to many assets, and terms unique to the asset at hand: a stock's alpha, a corporate bond's credit risk, or an option's implied volatility.

In this spirit, this paper proposes a model-based method to decompose the returns of AMM LP positions into two components. The first is a ``beta-like'' component, reflecting the \emph{market risk exposures} of the assets held by the AMM. We call this component of payoffs the \emph{rebalancing strategy}. The second, equal to the difference between the AMM's returns and the rebalancing strategy's returns, is an ``alpha-like'' component, which reflects \emph{microstructural} forces: accrued fees from noise traders minus adverse selection losses to arbitrageurs.

We develop our model in a simple continuous-time model. There is a risky asset and a num\'eraire, which can be traded on an AMM and an infinitely deep centralized exchange (CEX). As in the classic \citet{black1973pricing} setting, we assume the risky asset's price follows a geometric Brownian motion, with possibly stochastic volatility. Noise traders trade with the AMM, contributing fees to LPs. Arbitrageurs trade with the AMM and the CEX to maximize profits. We assume arbitrageurs pay no fees, implying that arbitrageurs ensure that the AMM's price is always equal to the CEX price. In the case that the AMM is a CFMM, the CFMM is described by an invariant curve $f\left(x,y\right)=L$; the CFMM is willing to make any trade such that it stays on this level curve.

Within our model, the profit and loss of the CFMM decomposes into two distinct components. The first reflects the AMM's role as a \emph{market maker}: LPs profit from any fees that accrue to the pool, and make adverse selection losses from arbitrageurs ``sniping'' the AMM's stale quotes when CEX prices move. Adverse selection losses admit a clean expression in our model, depending only on the instantaneous variance of asset prices, and the marginal liquidity of the AMM's current level set. We call this component ``alpha-like'', since it is unique to the microstructural tradeoffs facing the AMM.

The second component reflects the AMM's dual role as a \emph{mutual fund}. An AMM providing liquidity in the ETH-USDC holds a long position in ETH; it thus profits when ETH prices increase, and makes losses when ETH prices decrease, just as an actively managed mutual fund would. In our model, this component can be \emph{replicated} through an active trading strategy, which always holds the same amount of the risky asset as the AMM does, but makes all trades at CEX prices; we call this replication strategy the \emph{rebalancing strategy}. We call this component ``beta-like'', since it reflects simply directional exposure to underlying risky asset prices, rather than forces unique to the AMM as an investment.

In equity pricing models, stock betas are estimated by \emph{projecting} stock returns onto risk factors; the resultant residual defines the stock's alpha. Analogously, we show that the rebalancing strategy corresponds to the quadratic-variation-minimizing projection of AMM LP returns onto the risky asset's price movements, and loss-versus-rebalancing corresponds to the projection residual.

This perspective sheds light on the economic differences between the two return components. The ``beta-like'' rebalancing strategy is a risk-neutral martingale: since it simply holds a position in the risky asset, it only makes profits and losses to the extent that the risky asset has positive risk premia. In other words, the rebalancing strategy can be perfectly delta-hedged, leading its return to be identically zero. The ``alpha-like'' component is the difference between two strictly increasing processes: accrued fees from noise trade, and accrued losses to arbitrage traders. Just like stock alphas can be positive or negative, our alpha-like component can be greater or smaller than raw AMM LP returns. Either way, the alpha-like component captures the component of AMM returns that is not mechanically driven by beta exposure to the risky asset.

We use our model to empirically analyze the Uniswap v2 ETH-USDC trading pair. Implementing our procedure empirically is straightforward: the analyst only needs a time series of AMM holdings, mints, and burns, and a time series for the price of the risky asset. We construct discrete approximations to the continuous-time rebalancing strategy, which trade to match the AMM's risky asset holdings at frequencies between 1 minute and multiple months. We then consider how subtracting these strategies' profits from pool profits changes the volatility of the P\&L series.

The value of raw v2 ETH-USDC pool P\&L is fairly volatile: daily gains or losses of over \$10
million USD are common. However, the vast majority of this risk is simply market risk: subtracting
the profits of the rebalancing strategy reduces the variance of LP returns by four orders of
magnitude. Rebalancing frequencies between 1 minute and 1 hour produce similar results; volatility
is higher for longer periods between rebalancing, consistent with the idea that more market risk
accrues within-periods. Our model also implies that better approximations to the rebalancing
strategy should reduce P\&L volatility, not to zero, but to the limit of trading fees minus LVR;
this prediction holds in our data.

Our results suggest that, in empirical studies of AMM LP returns, it is quantitatively important
to hedge out market risk. Suppose an econometrician wishes to study the influence of some
$x$-variable on the microstructural component of AMM LP returns. In our analysis of the v2
ETH-USDC pair, 99.991\% of pool P\&L variance is driven by ``beta'' exposure to ETH prices; when
this component is subtracted, the remaining
``alpha'' component of fees minus adverse selection costs, which is presumably the target of the
econometrician's analyses, represents only 0.009\% of variance. In best case for the
econometrician, where market risk is simply orthogonal noise to the $x$-variables, using raw P\&L
as the $y$-variable would lead to around 108 times larger coefficient standard errors, so around
$108^2=11{,}664$ times more data are needed to achieve similar coefficient precision. Moreover,
even small correlations between $x$-variables and market risk can introduce omitted-variable bias
which overwhelms any real effect of $x$-variables on the AMM ``alphas''.

A common approach in industry practice is ``impermanent loss'', often defined as the difference between the current value of an AMM's reserves and the value of a static portfolio holding the same assets as the AMM at some fixed point in the past. Impermanent loss has a number of undesirable features as a loss metric. It does not aggregate cleanly over time: it can be positive on two short time intervals, but zero on the two intervals combined. Its value, and even its sign, can be manipulated through the analyst's arbitrary choice of start point.

Our approach fixes these issues simply by frequently updating the holdings benchmark: our
``alpha-like'' component is equivalent to dividing a long time interval into many small intervals,
calculating impermanent loss on each subinterval, and adding up the results. In mutual funds
terms, our approach can be thought of as calculating the fund's ``alpha'' on many small time
intervals, and adding up these alphas. Like standard return metrics in asset pricing, our loss
measure is additive and independent across time: its value in any period does not depend outcomes
in the distant past.

Finally, we discuss connections between AMM LP positions and the three classical ways that
volatility can be traded: static (European) options, dynamic trading strategies, and variance
swaps \citep{carr2001towards}. We model the AMM reserves at the static payoff of a pool value
function. This relates to \citet{clark2020replicating}, \citet{fukasawa2022weighted}, and
\citet{deng2023static}, who show that AMM LP payoffs, over any finite time horizon, can be
replicated by shorting a bundle of European options. These option positions, in turn, are
equivalent to dynamic trading strategies which sells (buys) the asset when prices increase
(decrease). In our setting, the rebalancing strategy plays the role of delta-hedging. Finally, a
delta-hedged LP position can be thought of as a generalized variance swap, whose payoff over any
given time period is equal to realized variance weighted by the marginal liquidity of the AMMs.

\subsection{Related Literature}

\paraheader{Empirical AMM Literature.} Our main contribution to the academic literature is to develop a model-based empirical approach for measuring AMM LP returns. We view this as filling a gap between the existing theoretical and empirical literatures on AMMs. There is a large and developed theoretical literature, mainly analyzing \emph{microstructural }models of AMMs \citep{aoyagi2020liquidity,aoyagi2021coexisting,capponi2021adoption,lehar2021decentralized,park2021conceptual,lehar2022liquidity,hasbrouck2022need,brolley2023demand}. Qualitatively, the microstructural forces in our paper are very similar to this literature, as well as classic microstructure models \citep{glosten1985bid}: AMM LPs benefit from noise trade, and suffer losses to informed arbitrageurs.

The majority of the aforementioned theoretical AMM papers are \emph{one-period} models; without dynamics, we will show that ``impermanent loss'' is unambiguously the right measure of AMM LP losses. But time is continuous in reality. In mapping the one-period AMM models to data, the empirical literature has faced a choice of whether to think of time as one long interval, or many shorter intervals. \citet{augustin2024reaching} takes the former approach, measuring losses relative to holdings at a fixed start date. \citet{aquilina2024decentralised} and \citet{fang2024learning} take the latter approach, measuring IL on a number of shorter intervals, and adding the results up. \citet{heimbach2022risks} do not specify whether the holdings benchmark is updated. All the above works claim to measure ``impermanent loss'', implying that the two approaches measure the same underlying economic object. There is currently no theoretical guidance on which measurement choice is correct, and how these decisions affect the measured economic outcomes.

In a simple dynamic model, we show that AMM LP returns contain both an ``alpha-like'' microstructural component, and a ``beta-like'' market risk component. Dynamics are necessary for this decomposition: we show in \Cref{sec:projection} that market risk is not meaningfully present in one-period models. We believe applied researchers should generally try to measure the alpha-like component: the beta-like component reflects the returns to a mechanical trading strategy, and is not specific to the AMM as an investment.

Our decomposition gives clear guidance on how the empirical literature should proceed. The ``one-interval'' and ``many-interval'' LP loss metrics in fact measure economically very different things. The ``many-interval'' metric, used in \citet{aquilina2024decentralised} and \citet{fang2024learning}, is exactly hedged LP returns, and thus converges to our ``alpha-like'' component of returns --- fees minus arbitrage costs. We think this is generally the right loss measure to use, though we recommend using a higher rebalacing frequency than that used in \citet{aquilina2024decentralised} and \citet{fang2024learning}. To avoid confusion, we suggest that researchers using this metric call it \emph{loss versus rebalancing} rather than impermanent loss.

In contrast, ``one-interval'' impermanent loss, as used in \citet{aquilina2024decentralised}, does not fully remove market risk from returns. The resultant metric combines microstructural forces and predictable exposure to risky asset prices, and our empirical analysis suggests that the latter can be much larger than the former. We recommend against using this metric, and we recommend that papers that choose to use this metric argue explicitly why it is meaningful to include market risk exposures in the definition of AMM LP returns, for the purposes of the question being studied.

\medskip{}
\paraheader{Theoretical AMM Literature.} While the microstructural economics of our model are very similar to the existing literature, we make a number of modelling simplifications in order to work cleanly in continuous time, which we discuss in detail in \Cref{subsec:assumptiondiscussion}. Some of these limitations have been addressed in follow-up work \citep{2023-02-lvr-fee-model}. Importantly, since our main modeling goal is to provide a framework for measuring LP returns, we purposefully do not impose any equilibrium model of the level of AMM liquidity in this paper: it is not possible to derive a measurement approach which allows detecting abnormal LP returns in a model which strongly assumes that LPs make normal returns. Ultimately, the goal of our model is to explain and motivate our empirical decomposition, which we view as our main contribution to the literature. The empirical approach can be used even if the reader does not agree with all the assumptions of our stylized model.

A number of papers outside the financial economics literature apply tools from convex analysis to analyze CFMMs. \citet{angeris2019analysis} show analytically and through agent-based simulations that CPMMs are able to closely track the reference prices of underlying assets. In a manner related to this paper, \citet{angeris2020improved} and \citet{angeris2021replicatingmarketmakers,angeris2021replicatingmonotonicpayoffs} apply tools from convex analysis (e.g., the pool reserve value function) to study the more general case of constant function market makers. \citet{angeris2021replicatingmonotonicpayoffs} also analyze LP profits, but do not relate them to the rebalancing strategy or perform measurements. Relative to these earlier papers, our contribution is to explicitly decompose LP losses into the returns on the rebalancing strategy, and obtained fees minus \LVR for general AMMs, to show how the \LVR component corresponds to arbitrageur profits, and to show how these ideas can be used to measure LP losses empirically.
Specifically, \citet{angeris2019analysis} only compute the \emph{expected} value under the risk
neutral distribution, only for \emph{constant product market makers}, deriving a dependence to
$\sigma^{2}/8$. We show, for \emph{all constant function market makers}, a \emph{path-by-path} decomposition of the LP returns (across \emph{all paths}) to the rebalancing strategy and characterize the leftover profit-and-loss which is noise trader fees minus \LVR.\footnote{In particular, a number of papers have characterized \emph{expected} losses of CFMM LPs---sometimes referred to as expected ``impermanent loss''---in more or less general cases.\citep{angeris2019analysis,evans2020liquidity}
As we discuss in Section \ref{sec:lvr-lvh}, the (risk-neutral) \emph{expectations} of ``impermanent loss'' and \LVR are equal, but \LVR does not contain market risk; we view this as an important conceptual distinction, and it also leads to cleaner empirical measures of LP losses.}

Subsequent to the initial draft of our paper, \citet{hasbrouck2023economic} build a model of equilibrium liquidity provision for concentrated-liquidity AMMs, using a continuous-time model with many of the same simplifying assumptions we use. The single-period and continuous-time modelling approaches are thus both used in the finance literature; the approaches each have their strengths and weaknesses, and we expect research to continue in both directions.

\medskip{}
\paraheader{Options.} Another group of papers draws analogies between CFMMs and options; we discuss the relationship between our results and these papers in detail in \Cref{sec:options}. \citet{clark2020replicating} replicates the payoff of a constant product market over a finite time horizon in terms of a static portfolio of European put and call options. \citet{tassy2020growth} compute the growth rate of a constant product market maker with fees. \citet{lambert2022} considers a number of related issues. \citet{boueri2021g3m} considers the profitability of geometric mean market makers under geometric Brownian motion dynamics; his (re-)definition of ``impermanent loss'' in that setting is equivalent to \LVR. \citet{boueri2021g3m} also does not show the relationship between \LVR and arbitrageur profits. Subsequent to the first publicly available version of our paper, \citet{cartea2022decentralised} and \citet{cartea2023predictable} discussed the concept of ``predictable loss''; the ``convexity cost'' component of predictable loss in their setting is equivalent to \LVR.

\medskip{}
\paraheader{Industry.} Since we introduced the concept, \LVR has evolved into a centerpiece in practical applications and analyses, having had wide impact within decentralized finance. \LVR and markouts have been used in practice to measure hedged LP returns of AMM pools, including those of concentrated liquidity market makers \citep{atis2023,crocswap_usage_2022}, and a number of industry groups are also working to build AMMs which utilize our insights to attempt to reduce or eliminate \LVR \citep{canidio2023arbitrageurs,catalabs2023,fenbushi2023,algebradexengine,gyroscope2023,aori,adams2024amm}.


\section{Institutional Background}
\label{sec:instbg}

Automated market makers have their origin in the classic literature on prediction markets and market scoring rules; see \citet{pennock2007computational} for a survey of this area. Constant function market makers, which are characterized by an invariant or bonding function, build on the utility-based market making framework of \citet{chenpennock2007}. In that framework, utility indifference conditions define a bonding function for binary payoff Arrow-Debreu securities. More recent interest in CFMMs has been prompted by an entirely new application: its functioning as a decentralized exchange mechanism, first proposed by \citet{vbuterin_lets_2016} and \citet{lu_building_2017}. 

An automated market maker (AMM) is a smart contract which allows market participants to trade one cryptoasset with another directly on the blockchain, rather than using a centralized exchange. Suppose, for example, that a market participant wanted to trade an asset such as ETH held in her blockchain wallet for another asset, such as USDC. She could do so on a custodial centralized exchange (CEX), such as Binance, by ``depositing'' her ETH by sending it to Binance on the blockchain; trading Binance-custodied ETH for custodied USDC; and then ``withdrawing'' the USDC, directing Binance to send USDC back into her personal blockchain wallet. This process has a number of costs. The market participant must give the CEX custody over her cryptoassets, exposing her to exchange credit risk. Market participants must have a CEX trading account; many CEXs impose jurisdictional and identification requirements, often imposed by national regulators to satisfy know-your-customer or anti-money-laundering laws, which market participants may be unwilling or unable to satisfy. The CEX must enable trading of the asset pair in question; many tokens with small market caps are simply not listed on CEXs for trade. Moreover, the rules for determining trade priority and prices on CEXs are not always transparent, and CEXs may impose deposit, withdrawal, and settlement fees or delays on their users. 

AMMs present an alternative to CEX trading which circumvents many of these costs. Technically, AMMs are ``smart contract'' wallets, meaning they are blockchain wallets which can hold cryptoassets, but whose behavior is determined fully by blockchain code rather than human discretion. In this example, the AMM wallet holds some inventory of ETH and USDC. When a trader submits a blockchain transaction to trade ETH for USDC, ETH is sent from the user to the AMM's inventory, and USDC is sent from the AMM's inventory to a user, in a single atomic transaction. We will discuss how the trade price is determined in Section \ref{sec:model} below. There are no barriers to access to AMMs: any individual can initiate a transaction with any AMM. There is no credit risk, since trades are completed in a single step, and it is not possible to lose the sold assets without gaining the purchased assets. The mechanisms through which trade prices are calculated is comparatively transparent: for example, the Uniswap~v2 smart contract consists of under a thousand lines of publicly available code.\footnote{See the \href{https://github.com/Uniswap/v2-core/tree/master/contracts}{Uniswap~v2 repository}.} Once ``deployed'' to the blockchain, no one, not even the creator, can modify the code. AMMs thus present an attractive trading option for market participants who do not want exposure to exchange credit risk, who face barriers to accessing CEXs, or who value the transparency of AMM behavior. 

The asset inventory AMMs use to trade is provided in a decentralized manner: any market participant can become a ``liquidity provider'' (LP) in a ETH-USDC AMM, or any other pair of tokens,\footnote{In particular, a market participant can begin LPing a pair of assets even if no one else is providing liquidity for it, thus effectively listing the pair of assets for trade.} by contributing ETH and USDC to the AMM's inventory.
The AMM pool then trades using these contributed assets as a part of its inventory, and any trading fees accrue proportionately to the liquidity inventory providers depending on her share. At any point, the liquidity providers can withdraw ETH and USDC corresponding to her share of the pool; however, the amounts of ETH and USDC withdrawn will generally differ from what the liquidity provider first contributed, as inventory adjusts with market participants' trade requests and fees are collected. The economic problem facing LPs, which is the core focus of our paper, is how to evaluate the costs and benefits of providing liquidity to AMMs.





\section{Model}
\label{sec:model}

\paraheader{Assets.} Fix a filtered probability space $\big(\Omega,\Fscr,\{\Fscr_{t}\}_{t\geq0},\Q\big)$ where $\Q$ is a risk-neutral or equivalent martingale measure, satisfying the usual assumptions. 
Suppose there are two assets,\footnote{This assumption is without loss of generality, we describe the multi-dimensional case where there are $n\geq2$ assets, none of which need be the num\'eraire, in \Cref{subsec:multidim}.} a risky asset $x$ and a num\'eraire asset $y$. We assume that the risk-free rate is zero.\footnote{This assumption is without loss of generality. In order to incorporate a non-zero risk-free rate, one need only add a cost of capital for the value of LP assets deposited as pool reserves.} There is an infinitely deep centralized exchange, where the risky asset can be traded with zero fees. The price on the centralized exchange is observable, and evolves exogenously according to a geometric Brownian motion that is a continuous $\Q$-martingale, i.e.,
\[
\frac{dP_{t}}{P_{t}}=\sigma_{t}\,dB_{t}^{\Q},\quad\forall\ t\geq0,
\]
with a stochastic volatility process\footnote{Volatility will be an important input in the analysis that follows. A natural question is how to calibrate volatility as a model parameter. As in the general application of Black-Scholes style models, for \emph{ex ante} analysis of a possible future LP position, an implied volatility is the appropriate input. On the other hand, for \emph{ex post} LP return performance analysis as in \Cref{sec:empirics}, a historical or realized volatility is appropriate.} given by $\sigma_{t}>0$, and where $B_{t}^{\Q}$ is a $\Q$-Brownian motion.

This is a standard set of assumptions in the mathematical finance literature, pioneered by
\citet{black1973pricing}, and accommodates assets with arbitrary risk premia: by Girsanov's
theorem, the risk-neutral measure simply removes the drift term from these assets' price
movements. Throughout the paper, we will mostly state our results using the risk-neutral measure
for expositional simplicity.\footnote{All of the key probabilistic results of the paper (e.g.,
  \Cref{lem:lvr},\Cref{th:decomposition}) hold both in the physical and risk-neutral probability
  measures. The only exception is part of \Cref{cor:lvb}, the implications of this are discussed in
  \Cref{fn:rnconttime}.}

\begin{figure}
\begin{subfigure}[t]{.475\textwidth} \centering \begin{tikzpicture}
      \begin{axis}[
        clip=false,
        no markers,
        axis lines=left,
        xtick=\empty,
        xticklabels={},
        ytick=\empty,
        yticklabels={},
        xlabel=$x$,
        ylabel=$y$,
        every axis y label/.style={at=(current axis.above origin),anchor=south},
        every axis x label/.style={at=(current axis.right of origin),anchor=west},
        small,
        xmin=0, xmax=3,
        ymin=0, ymax=3,
        ];
        \path[draw=black,dashed](axis cs:0.5,2) -- (axis cs:2,2)
        node [midway,below] {\small $x_1-x_0$};
        \path[draw=black,dashed] (axis cs:2,2) -- (axis cs:2,0.5)
        node [midway,right] {\small $y_0-y_1$};

        \addplot[mark=none,line width=1pt,domain=0.333:3] {1/x}
        node [right=2pt,pos=0] {\small $f(x,y) = L$};
        \path[draw=black,fill=purple] (axis cs:0.5,2) circle (1.5pt)
        node [above right] {\small $(x_0,y_0)$};
        \path[draw=black,fill=green] (axis cs:2,0.5) circle (1.5pt)
        node [below] {\small $(x_1,y_1)$};
      \end{axis}
    \end{tikzpicture} \caption{Transitions between any two points on the bonding curve $f(x,y)=L$ are permitted, if an agent contributes the difference into the pool.}\label{fig:cfmm-transition}
\end{subfigure} \hfill{}\begin{subfigure}[t]{.475\textwidth} \centering \begin{tikzpicture}
      \begin{axis}[
        clip=false,
        no markers,
        axis lines=left,
        xtick=\empty,
        xticklabels={},
        ytick=\empty,
        yticklabels={},
        xlabel=$x$,
        ylabel=$y$,
        every axis y label/.style={at=(current axis.above origin),anchor=south},
        every axis x label/.style={at=(current axis.right of origin),anchor=west},
        small,
        xmin=0, xmax=3,
        ymin=0, ymax=3,
        ];
        \addplot[mark=none,line width=1pt,domain=0.333:3] {1/x}
        node [right=2pt,pos=0] {\small $f(x,y) = L$};


        \addplot[domain=0.333:1.667,dashed] {2-x}
        node [below left,pos=0.9] {\small $\text{slope}=-P$};

        \path[draw=black,fill=orange] (axis cs:1,1) circle (1.5pt)
        node [above right] {\small $\big(x^*(P),y^*(P)\big)$};
      \end{axis}
    \end{tikzpicture} \caption{Arbitrageurs ensures that, when the price is $P$, pool reserves shift to the point on the bonding curve where the slope is equal to $-P$.}\label{fig:cfmm-price}
\end{subfigure} \caption{Illustration of a CFMM.}
\end{figure}
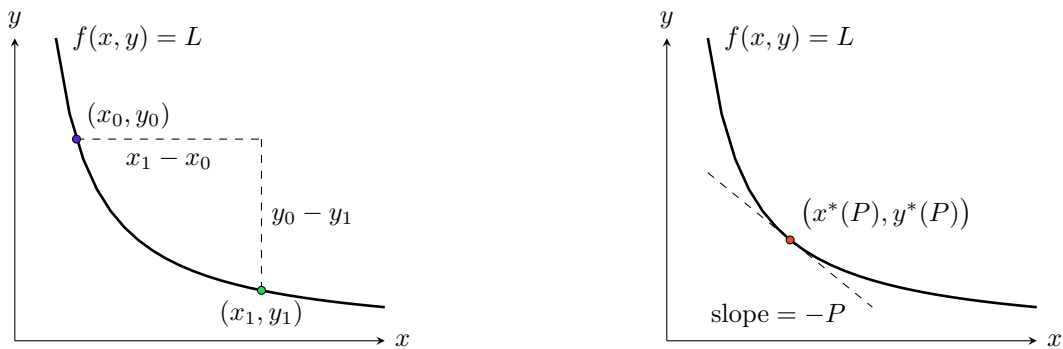

\medskip{}
\paraheader{CFMM pool.} The state of a constant function market maker (CFMM) pool is characterized by the reserves $(x,y)\in\R_{+}^{2}$, which describe the current holdings of the pool in terms of the risky asset and the num\'eraire, respectively. Define the feasible set of reserves \Cscr\ according to
\[
\Cscr\defeq\{(x,y)\in\R_{+}^{2}\ :\ f(x,y)=L\},
\]
where $f\colon\Rp^{2}\rightarrow\R$ is referred to as the \emph{bonding function} or \emph{invariant}, and $L\in\R$ is a constant. In other words, the feasible set is a level set of the bonding function. The pool is defined by a smart contract which allows an agent to transition the pool reserves from the current state $(x_{0},y_{0})\in\Cscr$ to any other point $(x_{1},y_{1})\in\Cscr$ in the feasible set, so long as the agent contributes the difference $(x_{1}-x_{0},y_{1}-y_{0})$ into the pool; see \Cref{fig:cfmm-transition}.

\begin{example} \label{ex:cpmm} The constant product market maker is defined by the invariant $\sqrt{xy}=L$. \end{example}

To simplify our analysis, we will also assume that, aside from trading with arriving liquidity demanding agents, the pool is static otherwise. In particular, we assume that the liquidity providers do not add (``mint'') or remove (``burn'') reserves over the time scale of our analysis. In other words, LPs are \emph{passive}. We also ignore the details of the underlying blockchain on which the pool operates. In particular, we assume away any blockchain transaction fees such as ``gas'' fees, and also ignore the discrete-time nature of block updates. %

Besides liquidity providers, there are two kinds of agents in the model: arbitrageurs and noise traders.

\medskip{}
\paraheader{Arbitrageurs.} There is a population of arbitrageurs, able to frictionlessly trade at the external market price, continuously monitoring the CFMM pool. When an arbitrageur interacts with the pool, we assume they maximize their immediate profit by exploiting any deviation from the external market price. In other words, they transfer the pool to a point in the feasible set $\Cscr$ that allows them to extract maximum value assuming that they unwind their trade at the external market price $P$. Geometrically, the presence of arbitrageurs implies that, if the price of the risky asset is $P$, pool reserves will move to the point on the curve $f\left(x,y\right)=L$ where the slope of the bonding curve is equal to $-P$, as indicated in \Cref{fig:cfmm-price}. %

\medskip{}
\paraheader{Noise traders.} There is also a population of DEX-specific noise traders. Noise traders trade only in the CFMM pool, and trade for totally idiosyncratic reasons. In most models of equilibrium liquidity provision on decentralized exchanges, it is important that there are at least some noise traders who strictly prefer trading on CFMMs to CEXs, since these traders are the source, since the equilibrium amount of CFMM liquidity is zero in the absence of noise traders \citep{lehar2021decentralized}. There are many reasons in practice why certain market participants might prefer trading on CFMMs to CEXs. Market participants may not be able or willing to satisfy the know-your-customer requirements imposed by CEXs, or may not trust that centralized exchanges will safely custody their assets. Market participants may, due to jurisdictional restrictions, not have access to a CEX on which a particular asset trades, whereas DEXes have no jurisdictional restrictions; market participants may also value the ability to atomically combine DEX trades with other smart contract operations on blockchains.\footnote{There could also be noise trading on the CEX; however, since we have assumed the CEX is infinitely deep, these trades do not affect prices and can be ignored in our model. If there are noise traders who can trade on either the CEX or the DEX, in our model they will choose to trade on the infinitely deep CEX. For the purposes of our model, we only need that there are some noise traders who exogenously prefer to trade on the DEX, with strong enough preferences that they are willing to use the DEX despite higher fees and price impact.}

Noise traders' trades have an initial impact on CFMM pool prices, but these effects are immediately offset by arbitrageurs, who immediately move the CFMM back to the CEX price $P$. Thus, from the LP's perspective, noise traders simply contribute a flow of fees. Denote by $\FEE_{T}$ the cumulative fees paid by noise traders up to time $T$. 
For simplicity, we assume fees are paid in units of the num\'eraire; this simplifies the analysis, since fees do not affect the level curve that the CFMM trades on.\footnote{The same assumption is made by \citet{lehar2021decentralized}.} In practice, fees are sometimes (e.g., in Uniswap~v2 but not in Uniswap~v3) reinvested into the pool reserves; another way to think about this assumption is to assume LPs immediately withdraw any accrued fees from the CFMM.

When discussing arbitrageurs, we will distinguish between two conceptually different forms of arbitrage activity. The first, which we call \emph{rebalancing arbitrage}, is arbitrage of a pool when mispricing arises due to movements of the CEX price. The second, which we call \emph{reversion arbitrage}, is arbitrage following the arrival of noise traders who move DEX prices away from $P$ --- this type of arbitrage is sometimes called ``back-running''. Our model allows us to quantify the magnitude of profits of rebalancing arbitrageurs, but not reversion arbitrageurs.

\subsection{Discussion of Assumptions}

\label{subsec:assumptiondiscussion}

The basic microstructural components in our model are qualitatively no different from the recent microstructure literature on AMMs \citep{lehar2021decentralized,capponi2021adoption,barbon2021quality,hasbrouck2022need,foley2023better} and classic microstructure models \citep{glosten1985bid}. As in the literature, AMM LPs earn fees from noise traders --- captured in reduced form by the cumulative fee process $\FEE_{t}$ --- and incur losses from trades with arbitrageurs.

The key difference between our model and the literature is that we work in continuous-time, whereas most of the microstructure literature uses single-period models. We do this because \emph{market risk} only arises meaningfully in dynamic models. In other words, one-period models suffice to analyze a CFMM's function as a \emph{market maker}, but its role as a \emph{mutual fund} only becomes evident in multi-period models. We work in continuous-time because it leads to simpler expressions than multi-period discrete-time models. However, in Subsection~\ref{subsec:binomialtree}, we show that our model's basic intuitions also hold in a two-step binomial tree.

Our goal is not to construct an \emph{equilibrium model of AMM liquidity provision}, as in much of the recent microstructure literature. Instead, our goal is to build a simple theoretical framework in which LP \pnl decomposes into a ``beta-like'' market risk component, and an ``alpha-like'' microstructural component. Two of our inputs --- the liquidity level $L$ and the fee process $\FEE_{t}$ --- are typically endogeneous in equilibrium models; we instead treat them as primitives. This is because, together with the bonding curve $f(x,y)$ and the volatility process $\sigma_{t}$, these inputs are sufficient statistics for determining the ``alpha-like'' component of returns. Any two equilibrium models that generate the same $L$ and $\FEE_{t}$ have the same implications for LP \pnl.

As an analogy, alphas in classical asset pricing settings capture deviations from frictionless equilibrium models. These alphas may have a variety of drivers, such as new risk factors, bounded rationality, or luck; the measurement approach for alphas does not take a direct stance on the drivers of alphas. Similarly, we provide a framework for measuring LP ``alphas'' which does not take a stance on the equilibrium models that drive these alphas. Our empirical methodology is thus valid across a wide range of equilibrium models.\footnote{While our primary goal is measurement, our model elements can be used to construct an equilibrium model of AMM liquidity. For example, \citet{hasbrouck2023economic} derives a stationary equilibrium liquidity distribution in concentrated-liquidity AMMs by solving for $L$ values that yield zero risk-neutral expected profits for LPs, net of fees. It is also in principle possible to design \pnl metrics which compare fees and arbitrage losses in a way incorporate optimizing LP behavior: for example, if the econometrician is willing to make the strong assumption that $\FEE_{t}$ is independent of $L$, the econometrician could calculate some liquidity level $L^{*}$ such that fees equal arbitrage losses, and use $L-L^{*}$ or $\frac{L}{L^{*}}$ as a loss metric. Most such metrics should reduce to some way to compare accrued fees and arbitrage losses. Different metrics may be more natural for different settings, and we do not pursue such approaches further in this paper.}

We assume the CEX is infinitely deep, so trades have no price impact. It is not entirely clear how true this is in practice, and for which assets this is true. On the one hand, \citet{barbon2021quality} and \citet{liao2022dominance} argue that transaction fees are in fact lower on DEXs than CEXs at certain transaction sizes. Moreover, since the barriers to listing tokens on DEXs are so low, many less liquid tokens begin trading on DEXs before they are traded on large CEXs. On the other hand, a large majority of trade volume in the USD-ETH pair, around 90\%, occurs on centralized exchanges relative to decentralized exchanges. Moreover, according to our conversations with industry participants, large orders are in practice often executed through OTC desks, so liquidity may be deeper than what is implied by CEX limit order books; the CEX in our model can be thought of as capturing limit order books, OTC desks, and any other sources of liquidity outside the DEX, including in fact other DEXes. Our model is a reasonable approximation so long as any individual DEX does not constitute more than a small fraction of total liquidity in the market available for an asset. Even in settings where this assumption does not hold, our model may still be a useful conceptual benchmark that is cross-sectionally consistent and arbitrage-free, analogous to the use of option pricing models to value options with illiquid underlying assets, such as employee stock options in privately held companies, or as in real options valuation \citep{dixit1994investment}.

We assume away many frictions to trading which are present in practice: we assume arbitrageurs pay no trading fees on CEXs or DEXs, we ignore gas fees, and we ignore the discrete, block-based nature of trading on blockchain CFMMs. Accounting for fees will imply that the profits arbitrageurs make will tend to be lower than our expressions. In particular, in practice, arbitrageurs tend to engage in ``gas races'', bidding high gas fees so that block miners have an incentive to include their arbitrage trades in the blockchain first. These gas races will tend to redirect some of the profits from CEX-DEX arbitrage towards block miners.\footnote{CEX-DEX arbitrage is one form of ``miner extractable value'', or MEV, a set of circumstances in which miners' ability to determine the ordering of transactions allows them to extract monetary value; \citet{daian2020flash} discusses MEV in detail.} 
The analysis of arbitrage profits in the presence of fees is the subject of follow-on work \citep{2023-02-lvr-fee-model}, and we defer a more careful discussion of the impact of fees to that work. We also assume noise trader fees are paid in the num\'eraire, and we assume away minting and burning of LP shares for simplicity. However, we relax both these assumptions in the empirical application in \Cref{sec:empirics}.

Finally, our stylized model is intended to motivate the decomposition and accompanying empirical methodology, not to serve as a fully realistic representation of AMM trading. This approach follows the tradition of classical asset pricing models like the CAPM \citep{sharpe1964capital,Lintner1965} and APT \citep{ross1976arbitrage}, whose simplifying assumptions motivate tractable empirical tools, despite their lack of literal realism. Our model yields a clean analytical decomposition that clarifies why LP returns can be viewed as the sum of a market-risk component and a microstructural component---two forces that should be separated empirically. Researchers need not accept all of the model\textquoteright s assumptions to apply the measurement framework it motivates.


\section{Market Risk ``Beta'' and Microstructural ``Alpha''}

\label{sec:analysis}

\paraheader{The pool value function $V\left(P\right)$.} \Cref{fig:cfmm-price} shows that the composition of the CFMM's reserve pool depends only on the risky asset's price: at any time $t$, if the risky asset's price is $P_{t}$, arbitrageurs will move the pool's reserves to the point on the $f\left(x,y\right)$ curve where the slope is $-P_{t}$. The mark-to-market value of the pool's reserves at any point in time, $P_{t}x_{t}+y_{t}$, is thus also fully determined by the current price $P_{t}$. A convenient way to analyze the monetary value of pool reserves at any given point in time is to define the \emph{pool value function} $V\colon\R_{+}\rightarrow\R_{+}$, as the solution to the optimization problem:
\begin{equation}
\begin{array}{lll}
V(P)\defeq & \minimize_{(x,y)\in\R_{+}^{2}} & Px+y\\
 & \subjectto & f(x,y)=L.
\end{array}\label{eq:pool-min}
\end{equation}
Intuitively, at any price $P$, arbitrageurs maximize profits by choosing the point on the curve $f(x,y)=L$ which minimizes the value of the pool's reserves. This minimizing choice is the tangency point illustrated in \Cref{fig:cfmm-price}; $V\left(P\right)$ measures the monetary value of reserves, $Px+y$, at this point. Hence, $V_{t}=V\left(P_{t}\right)$ at all times.\footnote{Note also that the optimization problem in \eqref{eq:pool-min} is isomorphic to the \emph{expenditure minimization problem }from classical demand theory: under price $P$, a consumer minimizes total expenditures $Px+y$, subject to staying on the indifference curve $f(x,y)=L$. The solution to this problem is to choose the point where the level curve of $f\left(x,y\right)$ is tangent to the budget set. The envelope theorem thus gives that the first derivative of the expenditure function is the Hicksian demand function, which is isomorphic to $x^{*}(P)$, and the second derivative of the expenditure function is the slope of Hicksian demand.}

We assume that the pool value function satisfies: \begin{assumption}\label{as:smooth-v}
\begin{enumerate}
\item \label{pt:opt} An optimal solution $\big(x^{*}(P),y^{*}(P)\big)$ to the pool value optimization \eqref{eq:pool-min} exists for every $P\geq0$.
\item \label{pt:smooth-v} The pool value function $V(\cdot)$ is everywhere twice continuously differentiable.
\item \label{pt:bounded} For all $t\geq0$,
\[
\E^{\Q}\left[\int_{0}^{t}x^{*}(P_{s})^{2}\sigma_{s}^{2}P_{s}^{2}\,ds\right]<\infty.
\]
\end{enumerate}
\end{assumption}

Parts~\ref{pt:opt}--\ref{pt:smooth-v} are easily verified for many CFMMs, see \Cref{sec:examples} for examples. Part~\ref{pt:bounded} is a square-integrability condition that will be used in \Cref{sec:analysis}. Parts~\ref{pt:opt}--\ref{pt:smooth-v} are a sufficient condition for the following: \begin{lemma}\label{le:envelope} For all prices $P\geq0$, the pool value function satisfies:
\begin{enumerate}
\item \label{pt:V} $V(P)\geq0$.
\item \label{pt:Vp} $V'(P)=x^{*}(P)\geq0$.
\item \label{pt:Vpp} $V''(P)=x^{*\prime}(P)\leq0$.
\end{enumerate}
\end{lemma}

Part~\ref{pt:Vp} of \Cref{le:envelope} establishes that the slope of the pool value function is equal to the reserves in the risky asset. Part~\ref{pt:Vpp} establishes that the pool value function is concave. Note that this concavity does not depend on the nature of the feasible set $\Cscr$ or the bonding function $f(\cdot)$. This part also establishes that the second derivative of the pool value function is the marginal liquidity available at the price level.

We call $x^{*}(P)$ the demand curve of the AMM. We remark that the concavity of the pool value function along with Part~\ref{pt:Vpp} of \Cref{le:envelope} establish that the demand curves of all CFMMs must be non-increasing functions of the implied pool price. This fact has deep economic roots, in that it is fundamentally implied by and equivalent to Myerson's lemma when LPs exchanging a risky asset with traders is viewed as an auction for which incentive compatibility of traders reporting their true valuations for the asset is required; this has been the subject of later investigations by \citet{milionis2023myersonian,jason_exchange_complexity}, and the monotonicity of demand curves holds for and forms the basis of a broader class of AMMs that are more general than CFMMs, and also include limit order books.

\begin{remark} The setting we accommodate and rely on is that of a locally-smooth demand curve $x^{*}(P)$. It is neither important nor required that this format comes from a CFMM, i.e., it is not necessary for our results and \LVR in particular that there is an invariant curve. So long as for some (instantaneous) interval of time, the LP is committing to a fixed demand curve according to \Cref{as:smooth-v}, the definition and characterization of \LVR in \Cref{lem:lvr} carries through precisely as-is with no modification. For one example, this remark means that the \LVR calculation of \Cref{lem:lvr} can be applied with absolutely no adjustments in general concentrated liquidity AMMs, such as Uniswap~v3 (in fact, see \Cref{ex:univ3} and \Cref{ex:univ3-pool}, where this is analyzed). This means that the concept of loss-versus-rebalancing, as we present it hereby, applies for \emph{any AMM} with a locally-smooth demand curve $x^{*}(P)$. Therefore, \LVR does not only refer to and is not a quantity specific to CFMMs; \LVR characterizes the behavior of general AMMs; also refer to \Cref{non-locally-smooth-footnote} for how to apply this to even more general, non-locally-smooth settings. \end{remark}

The pool value function allows us to write the profit and loss of an CFMM, from time $0$ to time $t$, as:
\begin{equation}
\text{LP \pnl}_{t}=V_{t}-V_{0}+\FEE_{t}.\label{eq:PandL}
\end{equation}
In words, $\text{LP \pnl}_{t}$ is the monetary value of the pool reserves at time $t$, minus the value at time $0$, plus the cumulative fees collected until time $t$.

\medskip{}
\paraheader{Rebalancing strategy $R_{t}$.} A key component of our decomposition is an object we call the \emph{rebalancing strategy.} Informally, the rebalancing strategy aims to always match the risky asset holdings of the CFMM. Whenever prices move and the CFMM changes its holdings of the risky asset, the rebalancing strategy makes exactly the same net trades; however, it executes these trades at CEX prices, rather than CFMM prices. The rebalancing strategy thus always has the same exposures to risky asset prices as the CFMM, and thus its \pnl will correspond to the ``market risk'' component of our decomposition.

Formally, we first define general trading strategies. A trading strategy is a process $(x_{t},y_{t})$ defining holdings in the risky asset and num\'eraire at each time $t$. For a trading strategy to be \emph{admissible}, we require that it be adapted, predictable, and satisfy
\begin{equation}
\E^{\Q}\left[\int_{0}^{t}x_{s}^{2}\sigma_{s}^{2}P_{s}^{2}\,ds\right]<\infty,\quad\forall\ t\geq0.\label{eq:x-bounded}
\end{equation}
We further restrict admissible trading strategies to be \emph{self-financing}, i.e., to satisfy
\begin{equation}
\underbrace{x_{t}P_{t}+y_{t}-\left(x_{0}P_{0}+y_{0}\right)}_{\pnl_{t}}=\int_{0}^{t}x_{s}\,dP_{s},\quad\forall\ t\geq0.\label{eq:self-fin}
\end{equation}
Self-financing trading strategies can be thought of as strategies which always adjust risky asset positions at market prices. Thus, the profits of any self-financing strategy are unaffected by its \emph{trades} --- arbitrary adjustments in $x_{t}$ do not directly impact profits. Instead, the profits of these strategies are entirely driven by their \emph{market risk exposures}: \Cref{eq:self-fin} states that the instantaneous change in a self-financing strategy's profits is equal to the risky asset position, $x_{s}$, times the change in price, $dP_{s}$. This also implies that, if we specify $y_{0}$, the initial amount of the num\'eraire, and $\{x_{t}\}_{t\geq0}$, risky asset holdings in all future histories, the path of the num\'eraire $\{y_{t}\}$ is implicitly determined through \eqref{eq:self-fin}. Note that, since $P_{t}$ is a \Q-martingale, the \pnl process given by \eqref{eq:self-fin} is also \Q-martingale, and by \eqref{eq:x-bounded} it is square-integrable.\footnote{When the risky asset has no risk premia, $P_{t}$ is also a physical-probability martingale, so self-financing trading strategies make zero profits in expectation, and conditional on any possible history. In the general case where $\mathbb{Q}\neq\mathbb{P}$ and the asset may have risk premia, the instantaneous expected return on any self-financing strategy is fully driven by the asset's risk premium, and any self-financing trading strategy makes zero profits when delta-hedged.}

We then define the rebalancing strategy to be the self-financing trading that starts initially holding $\big(x^{*}(P_{0}),y^{*}(P_{0})\big)$ (the same position as the CFMM), and continuously and frictionlessly rebalances to maintain a position in the risky asset given by $x_{t}\defeq x^{*}(P_{t})$. Let $R_{t}=P_{t}x_{t}+y_{t}$ denote the monetary value of the rebalancing strategy at time $t$; \eqref{eq:self-fin} implies that:
\begin{equation}
R_{t} - V_{0} = \int_{0}^{t}x^{*}(P_{s})\,dP_{s},\quad\forall\ t\geq0.\label{eq:rebalance}
\end{equation}
Because of \Cref{as:smooth-v} Part~\ref{pt:bounded}, the rebalancing strategy is admissible and $R_{t}$ is a square-integrable \Q-martingale. In particular, being a self-financing strategy, the rebalancing strategy breaks even in expectation under the risk-neutral measure \Q, and only makes expected returns to the extent that the underlying risky asset has nonzero risk premia.


We then express the change in pool value from time $0$ to time $t$ as the sum of the rebalancing strategy's profits, and a residual term which we will define as \emph{loss-versus-rebalancing}:
\[
V_{t}-V_{0}=\left(R_{t}-V_{0}\right)-\LVR_{t},
\]
\begin{equation}
\LVR_{t}\defeq R_{t}-V_{t}.\label{eq:lvrdef}
\end{equation}
$\LVR_{t}$ turns out to admit a simple interpretation as the losses of CFMMs to arbitrageurs from price slippage.

\begin{lemma}\label{lem:lvr} Loss-versus-rebalancing takes the form:
\begin{equation}
\LVR_{t}=\int_{0}^{t}\ell(\sigma_{s},P_{s})\,ds,\quad\forall\ t\geq0,\label{eq:lvr}
\end{equation}
where we define, for $P\geq0$, the instantaneous \LVR by:
\begin{equation}
\ell(\sigma,P)\defeq\frac{\sigma^{2}P^{2}}{2}\left|x^{*\prime}\left(P\right)\right|\geq0.\label{eq:ilvr}
\end{equation}
$\ell(\sigma,P)$ is always positive, so \LVR is a non-negative, non-decreasing, and predictable process. Moreover, the cumulative profits of rebalancing arbitrageurs up to time $t$ are equal to $\LVR_{t}$. \end{lemma}

\medskip{}
The core intuition behind \Cref{lem:lvr} is that the CFMM systematically loses money relative to the rebalancing strategy due to \emph{price slippage}: every trade made by the CFMM is made at slightly worse prices than the rebalancing strategy. \LVR is large when volatility $\sigma_{t}$ is high, so prices move more; and when $|x^{*\prime}\left(P_{t}\right)|$ is large, so the CFMM trades more of the risky asset when prices move a given amount. We will refer to $|x^{*\prime}\left(P_{t}\right)|$ as the \emph{marginal liquidity} of the CFMM at price $P_{t}$.\footnote{In \Cref{subsec:dxdpformula}, we derive an expression for $|x^{*\prime}\left(P\right)|$ in terms of the CFMM bonding function $f(\cdot)$, and show that $|x^{*\prime}\left(P\right)|$ is related to the curvature of the level sets of $f(\cdot)$: CFMMs with ``flatter'' bonding curves have higher marginal liquidity. An interesting implication of these results for the design of CFMM invariants is that, in our model, under our assumptions, the only feature of CFMMs which matters for losses is the curvature of the CFMM bonding function, which determines $x^{*\prime}\left(P\right)$. Any two CFMM invariants for an asset pair which have the same local convexity at any given price have the same $\LVR$.} The idea that CFMM LP losses are related to price volatility and bonding function curvature is well-understood in the theoretical literature on AMMs \citep{aoyagi2020liquidity,aoyagi2021coexisting,lehar2021decentralized,capponi2021adoption,oneill2022}; the relationship between these quantities is particularly simple in continuous-time models \citep{evans2020liquidity,angeris2020improved,angeris2021replicatingmarketmakers,angeris2021replicatingmonotonicpayoffs}.

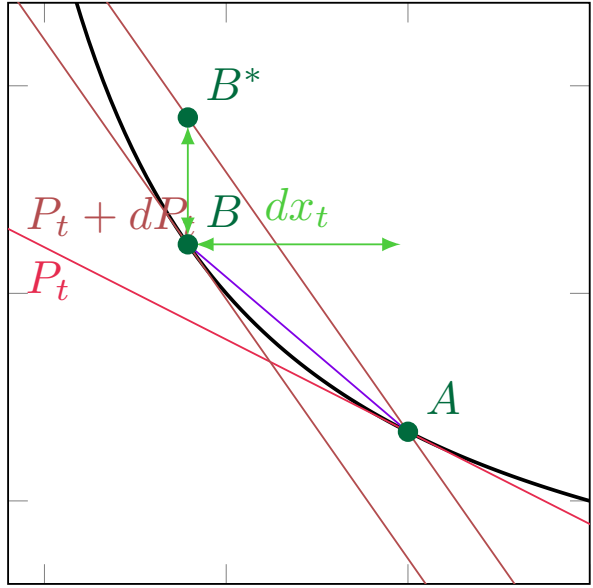
\begin{figure}
\global\long\def\arrowsep{1.5}%
\global\long\def\scaleparam{1.75}%
\global\long\def\textsizemult{1.5}%

\begin{centering}
\begin{tikzpicture}[scale = \scaleparam]

\definecolor{utilcolor}{rgb}{0.0, 0.0, 1.0};
\definecolor{pricecolor1}{rgb}{0.9, 0.17, 0.31};
\definecolor{pricecolor2}{rgb}{0.7, 0.3, 0.31};
\definecolor{pricecolor3}{rgb}{0.5, 0, 0.9};
\definecolor{pointcolor}{rgb}{0.0, 0.42, 0.24};
\definecolor{gapcolor}{rgb}{0.3, 0.8, 0.24};

\tikzstyle{l} = [draw, -latex', ultra thick]
\tikzstyle{block} = [rectangle,
        draw=utilcolor,
        thick,
        fill=utilcolor!20,
        text width=5em,
        align=center,
        rounded corners,
        minimum height=2em,
        scale = \scaleparam*\textsizemult]

\tikzstyle{l} = [draw, -latex',ultra thick]

\tikzmath{
\myres = 500;
\XEM0 = 1.5;
\YEM0 = 0.6666;
\p0 = 0.444444;
\XEM1 = 0.8944267;
\YEM1 = 1.118033;
\YRB1 = 1.423633;
\p1 = 1.25;
\greenlineX1 = \XEM0 - 0.02;
\greenlineX2 = \XEM1 + 0.02;
\greenlineXmid = (\XEM0 + \XEM1) / 2;
\XEMmid = (\XEM0 + \XEM1) / 2;
\YEMmid = (\YEM0 + \YEM1) / 2;
\YEMRBMID = (\YEM1 + \YRB1) / 2;
}

\begin{axis}[
width = 6cm, height = 6cm,
xmin = 0.4, xmax = 2,
ymin = 0.3, ymax = 1.7,
xtick = {}, ytick = {},
xticklabels={,,},
yticklabels={,,}
]
\addplot[thick, domain=0:10,samples=500] {1/x};

\addplot[pointcolor, mark = *, only marks]
coordinates{
(\XEM0,\YEM0) 
(\XEM1, \YEM1) 
(\XEM1, \YRB1) 
};

\node[pointcolor] at (axis cs:\XEM0,\YEM0) [anchor=south west] {\small $A$};
\node[pointcolor] at (axis cs:\XEM1, \YEM1) [anchor=south west] {\small $B$};
\node[pointcolor] at (axis cs:\XEM1, \YRB1) [anchor=south west] {\small $B^*$};

\node[pricecolor1] at (axis cs:0.4,1.03) [anchor=west] {\small$P_{t}$};
\addplot[pricecolor1, domain=0:10,samples=\myres] {\YEM0+(-\p0)*(x-\XEM0)};

\node[pricecolor2] at (axis cs:0.4,1.185) [anchor=west] {\small$P_{t}+dP_{t}$};
\addplot[pricecolor2,domain=0:10,samples=\myres] {\YEM0+(-\p1) * (x-\XEM0)};
\addplot[pricecolor2,domain=0:10,samples=\myres] {\YEM1+(-\p1) * (x-\XEM1)};

\draw[pricecolor3] (axis cs:\XEM0,\YEM0) -- (axis cs:\XEM1, \YEM1);


\draw[latex-latex,gapcolor] (axis cs:\XEM1,\YEM1+0.02) -- (axis cs:\XEM1, \YRB1-0.02);

\draw[latex-latex,gapcolor] (axis cs:\greenlineX1,\YEM1) -- (axis cs:\greenlineX2,\YEM1);
\node[gapcolor] at (axis cs:\greenlineXmid,\YEM1) [anchor=south] {\small$dx_{t}$};

\end{axis}







\end{tikzpicture}
\par\end{centering}
\caption{\LVR and a stylized depiction of CFMM LP price slippage. Suppose prices begin at $P_{t}$, the slope of the red line; the CFMM reserves then begin at point $A$. If prices increase to $P_{t}+dP_{t}$, the slope of the brown line, the CFMM trades to point $B$. The rebalancing strategy trades instead at the price $P_{t}+dP_{t}$, to point $B^{*}$. \LVR is the vertical gap between $B$ and $B^{*}$.}\label{fig:slippage}
\end{figure}

We prove \Cref{lem:lvr} rigorously in \Cref{sec:proofs}; here, we present an intuitive derivation based on \Cref{fig:slippage}. Suppose the market price changes from $P_{t}$ to $P_{t}+dP_{t}$. Arbitrageurs thus trade with the CFMM, moving from point $A$ to point $B$ on the CFMM invariant curve, and selling $dx_{t}$ of the risky asset, indicated by the horizontal green line. The rebalancing strategy sells exactly the same amount $dx_{t}$, but trades at the CEX price, $P_{t}+dP_{t}$; it trades along the brown line of slope $P_{t}+dP_{t}$ passing through $A$, reaching point $B^{*}$, which is higher than point $B$. Thus, after the price change, the CFMM and the rebalancing strategy hold the same amount of the risky asset, but the rebalancing strategy holds $B^{*}-B$ more cash than the CFMM.

The rebalancing strategy trades at the slope of the brown line, $P_{t}+dP_{t}$. The CFMM trades at the slope of the purple line --- that is, the secant line connecting points $A$ and $B$. The tangency lines at points $A$ and $B$ have slopes $P_{t}$ and $P_{t}+dP_{t}$ respectively, so the secant line has slope $P_{t}+\frac{dP_{t}}{2}$. Thus, the $B-B^{*}$ line has height:
\begin{equation}
dx_{t}\times\left(\left(P_{t}+dP_{t}\right)-\left(P_{t}+\frac{dP_{t}}{2}\right)\right)=\frac{dx_{t}\,dP_{t}}{2}.\label{eq:LPloss}
\end{equation}
This is thus the CFMM's cash loss relative to the rebalancing strategy. It is also equal to arbitrageur profits: arbitrageurs purchase quantity $dx_{t}$ from the CFMM at $P_{t}+\frac{dP_{t}}{2}$, and sell to the CEX at $P_{t}+dP_{t}$, earning $\frac{dx_{t}dP_{t}}{2}$.

Next, we write the amount traded $dx_{t}$ as a function of $dP_{t}$ and the slope of the CFMM's demand function, $x^{*}\left(P_{t}\right)$:
\begin{equation}
dx_{t}=\left|\frac{dx^{*}(P)}{dP}\right|\,dP_{t}=\left|x^{*\prime}\left(P_{t}\right)\right|\,dP_{t},\label{eq:dxdptemp}
\end{equation}
Expression \eqref{eq:LPloss} then becomes:
\begin{equation}
\frac{dx_{t}\,dP_{t}}{2}=\left|x^{*\prime}\left(P_{t}\right)\right|\,\frac{\left(dP_{t}\right)^{2}}{2}.\label{eq:lvrtemp1}
\end{equation}
Now, for a geometric Brownian motion, in a small amount of time $dt$, the quadratic variation $\left(dP_{t}\right)^{2}=d[P_{t}]$ is equal to $\sigma_{t}^{2}P_{t}^{2}\,dt$, that is, the instantaneous variance $\sigma_{t}^{2}\,dt$ multiplied by the square of the price. Plugging in to \eqref{eq:lvrtemp1}, the instantaneous losses of the CFMM LP position are equal to \eqref{eq:ilvr} of \Cref{lem:lvr}:
\begin{equation}
\left|x^{*\prime}\left(P_{t}\right)\right|\,\frac{\sigma_{t}^{2}P_{t}^{2}}{2}\,dt.\label{eq:lvrexptemp}
\end{equation}

The derivation illustrates that CFMMs rebalancing --- selling when prices rise and buying when prices fall --- is not in itself a source of losses. The rebalancing strategy makes exactly the same trades of the risky asset as the CFMM LP position, but does not lose money because it executes all trades at CEX prices.\footnote{For that matter,\emph{ any} trading strategy which executes all trades at CEX prices exactly breaks even under the risk-neutral measure.} $\LVR$ arises from price slippage: the fact that CFMMs execute all trades at worse-than-market prices.

The arbitrage losses of CFMMs are closely related to models of ``quote sniping'' in high-frequency trading. In \citet{budish2015high}, price changes render quotes ``stale'', triggering a race between market makers trying to cancel their quotes and arbitrageurs seeking to snipe them. CFMMs in our model can be thought of as market makers that never proactively update quotes: CFMM prices become stale whenever CEX prices move, and loss-versus-rebalancing captures the cumulative losses LPs suffer from arbitrageur ``quote sniping''.\footnote{We model LPs as fully passive. In practice, LPs could attempt to avoid arbitrage losses either by withdrawing liquidity upon large price movements, or attempting to ``snipe'' their own quotes before arbitrageurs can. In a frictionless model where LPs and arbitrageurs competed on equal footing whenever CEX prices moved, any possible arbitrage profits from sniping would be dissipated through gas fees; thus, we conjecture that \LVR would accurately capture LPs losses in such an environment, though it would not capture arbitrageur profits. CEX-DEX arbitrage is more complicated in reality; modelling this accurately is potentially helpful for understanding AMM behavior quantitatively, but is outside the scope of the current paper. }

The basic microstructural tradeoff facing the CFMM LP in our model is no different from \citet{budish2015high} and the classic microstructure literature: LPs gain from fees, which we model as an exogeneous increasing process, and lose from trades with arbitrageurs. A subtle difference between our model and that of \citet{budish2015high} and the classic literature is that losses in our setting are driven by \emph{volatility} --- knowledge of \emph{current} market prices, which only transmits into the CFMM's prices through arbitrage. \Cref{lem:lvr} shows that these losses can be quantified based on the volatility of CEX prices and the CFMM's mechanical price-quantity schedule. In contrast, \emph{informed trading} in classic microstructure models --- trading driven by knowledge of \emph{future} market prices --- is much more difficult to quantify.

We now state the main result of our model, which follows immediately from \Cref{eq:lvrdef} and \Cref{lem:lvr}.

\begin{theorem}\label{th:decomposition}We have:
\begin{equation}
\mathrm{LP\ \pnl}_{t}=\FEE_{t}+V_{t}-V_{0}=\underbrace{\int_{0}^{t}x^{*}(P_{s})\,dP_{s}}_{\text{market risk ``beta''}}+\underbrace{\FEE_{t}-\LVR_{t}}_{\text{microstructural ``alpha''}}\label{eq:PnLdecomp}
\end{equation}
\end{theorem}

\Cref{th:decomposition} states that $\text{LP \pnl}_{t}$ can be decomposed into a ``beta-like'' market risk component, and an ``alpha-like'' microstructural component.

The ``beta-like'' term reflects \emph{market risk}: directional exposure to the risky asset's price changes. Intuitively, the CFMM behaves like a rule-following actively managed mutual fund or ETF, mechanically adjusting its risky asset holdings $x^{*}(P)$ as prices change. The CFMM thus makes instantaneous profits equal to the product of its risky asset holdings, $x^{*}(P_{s})$, and instantaneous changes in risky asset prices, $dP_{s}$. Like market or factor betas in equity pricing settings, this component is not \emph{unique} to the CFMM; in our model, it can be attained simply by replicating the AMM's trades on the CEX, holding $x^{*}(P)$ when the price is $P$, as the rebalancing strategy does.

The ``alpha-like'' term reflects the \emph{microstructural} tradeoff facing the CFMM: as in
classic microstructure models \citep{glosten1985bid}, LPs make money from fee-paying noise trade,
and lose money to informed arbitrage. Like alphas in equity-pricing settings, this term is
comparatively \emph{unique} to the CFMM, in the sense that it cannot be replicated simply by
trading the underlying asset.\footnote{A caveat is that \Cref{lem:lvr} implies that $\LVR_{t}$ is
  mechanically related to realized variance, and thus can be replicated by trading instruments
  such as options and variance swaps; we discuss this in \Cref{sec:options}.
  By assumption in our model, $\FEE_{t}$ is not in general
  mechanically related to price movements in the underlying asset, and thus cannot in general be
  replicated by trading the underlying asset.}

In equity-pricing settings, a stock's alpha can be isolated by hedging exposures to market or factor risk, by subtracting the returns of ``tracking portfolios'' with zero alphas and equal betas to the stock. Analogously, we will show in Section \ref{sec:empirics} that the microstructural ``alpha'' component can be isolated simply by subtracting the profits of the rebalancing strategy from LP \pnl.


\section{The Rebalancing Strategy as a Projection}

\label{sec:projection}

In this section, we show that the rebalancing strategy corresponds to the quadratic-variation-minimizing \emph{projection} of AMM profits onto the risky asset's price path, analogous to how a stock's betas are calculated by projecting its returns onto market or factor returns. This gives a precise technical foundation for interpreting the rebalancing strategy as capturing the component of AMM \pnl ``explained by'' movements in the underlying asset's price. We first build intuition using a simple two-step binomial tree, then prove the result formally in continuous time.

\subsection{Binomial Tree}\label{subsec:binomialtree}

We first analyze a simple two-step binomial tree, illustrated in \Cref{fig:binomtree}. We consider a constant-product CFMM with invariant $f\left(x,y\right)=xy$, which begins holding $\left(x_{0},y_{0}\right)=\left(1,1\right)$. CEX price movements are depicted in Panel A. Prices begin at $P_{0}=1$; at $t=1$, they increase to $P_{1}^{U}=1.4$ or decrease to $P_{1}^{D}=0.6$; at $t=2$, they either revert to $P_{2}^{DU}=P_{2}^{UD}=1$, or diverge further to $P^{UU}=1.8$ or $P^{DD}=0.4$. The tree is constructed so that, under zero interest rates, risk-neutral probabilities of up and down movements are all 0.5; at these probabilities, prices are martingales, and any strategy that trades at market prices has zero expected return. We ignore noise trade and fees in the tree for expositional simplicity.

We can calculate the asset positions and profits of three strategies: \emph{buy-and-hold}, which simply holds the initial endowment $\left(x_{0},y_{0}\right)=\left(1,1\right)$ forever, in Panel B; the rebalancing strategy in Panel C; and the CFMM in Panel D. For the CFMM, we compute what $\left(x_{t},y_{t}\right)$ is as a function of $P_{t}$ and the initial state $\left(x_{0},y_{0}\right)$; for example, at price $P^{U}=1.4$, we must have $x_{0}y_{0}=1$ and $\frac{y_{0}}{x_{0}}=1.4$, implying $\left(x_{1},y_{1}\right)=\left(0.845,1.183\right)$. The rebalancing strategy simply mimics the CFMM's changes in $x_{t}$, but trades from $x_{t}$ to $x_{t+1}$ at price $P_{t+1}$, so its cash position updates as:
\[
y_{t+1}=y_{t}-P_{t+1}\left(x_{t+1}-x_{t}\right)
\]
The figure shows the asset holdings of each strategy at each node, as well as the market value of holdings, $P_{t}x_{t}+y_{t}$.

We define \emph{loss versus holding }(\LVH)\emph{ }as the difference between the CFMM in panel D, and the buy-and-hold strategy in panel B. \emph{Loss versus rebalancing} is instead the difference between the CFMM in panel D and the rebalancing strategy in panel C. The difference between \LVH and \LVR --- the B-C payoff difference --- represents the payoffs of the rebalancing strategy relative to buy-and-hold.

The B-C difference is a \emph{risk-neutral martingale}: since it makes all trades at market prices, it breaks even in risk-neutral expectation at each node. At $t=1$, Panel B and C payoffs are identical, since rebalancing occurs after price movements. Payoffs diverge at $t=2$. On the upper path, the rebalancing strategy sells to $x_{1}^{U}=0.845$, holding less of the risky asset than buy-and-hold. This increases payoffs if prices revert to $P_{2}^{UD}$ (2.062 vs 2.000), and lowers them if prices diverge to $P_{2}^{UU}$ (2.738 vs 2.800). These gains and losses offset exactly: risk-neutral expected profits from $P_{1}^{U}$ onwards are 2.400 for both B and C. Similarly, on the downward path, the rebalancing strategy buys to $x_{1}^{D}=1.291$, outperforms at $P_{2}^{DU}$ and underperforms at $P_{2}^{DD}$, and earns 1.6 in expectation.

The C-D difference is a \emph{strictly increasing process.} C makes the same trades as D, so C and D have identical risky asset holdings $x_{t}$ in all tree notes. However, C makes trades at CEX prices, and D trades at the worse AMM prices. Thus, D always underperforms C, and the C-D gap is strictly increasing in $t$.

\Cref{fig:binomtree} thus shows that \LVH can be thought of as the sum of the B-C gap, a risk-neutral martingale, and the C-D gap, a strictly increasing process. Removing the B-C gap from \LVH does not affect risk-neutral expected losses. Rather, it is useful for \emph{denoising}: it takes out directional exposure to the risky asset's price, node by node, leaving only losses driven by price slippage.\footnote{\label{fn:rnbinom}Following tradition \citep{cox1979option}, we analyze the binomial tree under the risk-neutral measure for simplicity. An interpretation of our results, when the underlying asset may have risk premia and physical probabilities differ from risk-neutral probabilities, is that the expected returns of the buy-and-hold and rebalancing strategies are fully driven by their \emph{deltas} --- their linear exposures to the risky asset's price, which here is exactly their $x_{t}$ values. For example, if the underlying asset had positive risk premia, the rebalancing strategy would overperform (underperform) buy-and-hold from $P^{D}$ ($P^{U}$) onwards, purely because it holds more of the high-return risky asset; hedging this directional exposure would also eliminate the expected return difference. The risk-neutral measure thus allows us to distinguish between return differences driven by the underlying asset's risk premium, and residual return differences.}

A seemingly attractive feature of loss-versus-holding is that, if prices revert to $1$ --- as in states $P_{2}^{UD}$ and $P_{2}^{DU}$ --- \LVH is equal to zero. This is the basis of the colloquially popular idea that CFMM losses are ``impermanent'': losses relative to buy-and-hold vanish, as long as prices eventually revert. Figure \Cref{fig:binomtree} shows where this intuition is misleading. The trading strategy of the CFMM --- looking at the movements of $x_{t}$, common to Panels C and D --- is a bet on price convergence. Executed at market prices, as in the rebalancing strategy in Panel C, this trading strategy profits upon mean-reversion, and loses money upon divergence. The CFMM executes all trades at worse-than-market prices, so it is a very inefficient bet on convergence: it breaks even instead of profiting if prices mean-revert, and loses even more than the rebalancing strategy when prices diverge.

\begin{figure}

        \begin{center}
                \footnotesize
                \textbf{\sffamily Panel A: Binomial Tree}

                \begin{forest}
                        for tree={grow'=east,l sep=4em,s sep=3em,inner sep=2pt }
                        [{$P_0=1.0$}, align = center, base = bottom
                        [{$P^U_1=1.4$}
                        [{$P^{UU}_2=1.8$}]
                        [{$P^{UD}_2=1.0$}]
                        ]
                        [{$P^D_1=0.6$}
                        [{$P^{DU}_2=1.0$}]
                        [{$P^{DD}_2=0.2$}]
                        ]
                        ]
                \end{forest}

                \vspace{2ex}

                \textbf{Panel B: Buy and Hold Strategy}

                \begin{forest}
                        for tree={grow'=east,l sep=0.8em,s sep=0.8em,inner sep=0.8pt }
                        [{$\Pi_t=2.000$\\$(x_t,y_t)=(1.000,1.000)$}, align = center, base = bottom
                        [{$\Pi_t=2.400$\\$(x_t,y_t)=(1.000,1.000)$}, align = center, base = bottom
                        [{$\Pi_t=2.800$\\$(x_t,y_t)=(1.000,1.000)$}, align = center, base = bottom]
                        [{$\Pi_t=2.000$\\$(x_t,y_t)=(1.000,1.000)$}, align = center, base = bottom]
                        ]
                        [{$\Pi_t=1.600$\\$(x_t,y_t)=(1.000,1.000)$}, align = center, base = bottom
                        [{$\Pi_t=2.000$\\$(x_t,y_t)=(1.000,1.000)$}, align = center, base = bottom]
                        [{$\Pi_t=1.200$\\$(x_t,y_t)=(1.000,1.000)$}, align = center, base = bottom]
                        ]
                        ]
                \end{forest}

                \vspace{2ex}

                \textbf{Panel C: Rebalancing Strategy}

                \begin{forest}
                        for tree={grow'=east,l sep=0.8em,s sep=0.8em,inner sep=0.8pt }
                        [{$\Pi_t=2.000$\\$(x_t,y_t)=(1.00,1.00)$}, align = center, base = bottom
                        [{$\Pi_t=2.400$\\$(x_t,y_t)=(0.845,1.217)$}, align = center, base = bottom
                        [{$\Pi_t=2.738$\\$(x_t,y_t)=(0.756,1.378)$}, align = center, base = bottom]
                        [{$\Pi_t=2.062$\\$(x_t,y_t)=(1.014,1.048)$}, align = center, base = bottom]
                        ]
                        [{$\Pi_t=1.600$\\$(x_t,y_t)=(1.291,0.825)$}, align = center, base = bottom
                        [{$\Pi_t=2.116$\\$(x_t,y_t)=(1.032,1.084)$}, align = center, base = bottom]
                        [{$\Pi_t=1.084$\\$(x_t,y_t)=(2.308,0.622)$}, align = center, base = bottom]
                        ]
                        ]
                \end{forest}

                \vspace{2ex}

                \textbf{Panel D: CFMM LP}

                \begin{forest}
                        for tree={grow'=east,l sep=0.8em,s sep=0.8em,inner sep=0.8pt }
                        [{$\Pi_t=2.000$\\$(x_t,y_t)=(1,1)$}, align = center, base = bottom
                        [{$\Pi_t=2.366$\\$(x_t,y_t)=(0.845,1.183)$}, align = center, base = bottom
                        [{$\Pi_t=2.683$\\$(x_t,y_t)=(0.745,1.341)$}, align = center, base = bottom]
                        [{$\Pi_t=2.000$\\$(x_t,y_t)=(1.00,1.00)$}, align = center, base = bottom]
                        ]
                        [{$\Pi_t=1.549$\\$(x_t,y_t)=(1.291,0.775)$}, align = center, base = bottom
                        [{$\Pi_t=2.000$\\$(x_t,y_t)=(1.00,1.00)$}, align = center, base = bottom]
                        [{$\Pi_t=0.894$\\$(x_t,y_t)=(2.236,0.447)$}, align = center, base = bottom]
                        ]
                        ]
                \end{forest}
        \end{center}

        \caption{The performance of buy-and-hold,
                a constant-product CFMM, and the rebalancing strategy, on a two-step binomial
                tree. Panel A depicts the binomial tree. The performance of the buy-and-hold
                strategy is shown in Panel B; the rebalancing strategy is shown in
                panel C, and the constant product CFMM is shown in panel D. In each
                panel, $x_{t},y_{t}$ are the holdings of the strategy, and $\Pi_{t}$
                is the pool value, $y_{t}+P_{t}x_{t}$.\label{fig:binomtree}}

\end{figure}
\subsection{Continuous Time}

\label{subsec:conttimeprojection}

With some more technical machinery, we can derive a continuous-time equivalent of the decomposition suggested by \Cref{fig:binomtree}. We consider a broad class of \emph{benchmark strategies}, which begin holding the same position in the risky asset as the CFMM, and adjust holdings at CEX prices. Formally, we define a benchmark as a self-financing trading strategy, described by a position $\bar{x}_{t}$ in the risky asset. Initial holdings match the pool, i.e., $(\bar{x}_{0},\bar{y}_{0})\defeq\big(x^{*}(P_{0}),y^{*}(P_{0})\big)$. We assume that $\bar{x}_{t}$ satisfies the square-integrability condition \eqref{eq:x-bounded}, so that the resulting trading strategy is admissible.

Because it is self-financing, the only way that the benchmark strategy can make or lose money is
through it's exposure to the risky asset. Hence, the \emph{value} of the benchmark strategy,
$\bar{R}_{t}(\bar x)$, depends on the holding strategy $\bar{x}_{t}$ and movements in prices, according to \eqref{eq:self-fin}:
\[
\bar{R}_{t}(\bar x)=V_{0}+\int_{0}^{t}\bar{x}_{s}\,dP_{s},\quad\forall\ t\geq0.
\]
We define the AMM's loss versus this benchmark as $\LVB_{t}(\bar x)\defeq\bar{R}_{t}(\bar
x)-V_{t}$.

\begin{corollary} \label{cor:lvb} For all $t\geq0$,
\begin{equation}
  \begin{split}\LVB_{t}(\bar x) \defeq\bar{R}_{t}(\bar x)-V_{t}
    &
      =\LVR_{t}+\underbrace{\int_{0}^{t}\left[
      \bar{x}_{s}-x^{*}(P_{s})
      \right]\,dP_{s}}_{\defeq\Delta(\bar{x})_{t}},
  \end{split}
\label{eq:LVBeqn}
\end{equation}
and at the same time,
\begin{equation}\label{eq:LVBmean}
\E^{\Q}\left[\LVB_{t}(\bar x)\right]=\E^{\Q}\left[\LVR_{t}\right].
\end{equation}
The loss process has quadratic variation\footnote{\label{fn:rnrw}Note that \eqref{eq:LVBeqn} and \eqref{eq:LVBqv}
  hold almost surely over all sample paths under the risk-neutral measure. Hence, they also hold almost surely over all sample paths under the physical
  measure.}
\begin{equation}\label{eq:LVBqv}
[\LVB(\bar x)]_{t}=[\Delta(\bar{x})]_{t}=\int_{0}^{t}\left[\bar{x}_{s}-x^{*}(P_{s})\right]^{2}\,\sigma_{s}^{2}P_{s}^{2}\,ds\geq[\LVR]_{t}=0.
\end{equation}
Therefore, among all benchmark strategies, the rebalancing strategy uniquely (up to a set of times
of measure zero) defines a loss
process with minimal (zero) quadratic variation.
\end{corollary}
\begin{proof} The first part is
an immediate corollary of \Cref{lem:lvr} and \eqref{eq:rebalance}. The second part is immediate by
the definition of the risk-neutral measure. The third part follows from the It\^o
isometry. \end{proof}

\Cref{cor:lvb} characterizes the impact of the choice of benchmark self-financing trading
strategy. It establishes that the residual loss $\LVB(\bar x)$ to an arbitrary benchmark $\bar x$
that trades as market prices is given by $\LVR$ plus an additional term $\Delta(\bar x)$. By examining
the integrand in \eqref{eq:LVBqv}, it is clear that, on a path-by-path basis, the residual loss
process $\LVB(\bar x)$
must have positive quadratic variation, unless $\bar{x}_s = x^*(P_s)$ for almost all
$s \in [0,t]$. This condition characterizes the rebalancing
strategy.\footnote{\label{fn:rnconttime}The choice of benchmark does not affect risk-neutral
  \emph{expected} payoffs. All choices of $\bar{x}$ lead to benchmarks with equal risk-neutral
  \emph{expected} returns, cf.\ \eqref{eq:LVBmean}. Equivalently, any physical expected return
  differences are purely driven by asset risk premia, and disappear after
  delta-hedging. \eqref{eq:LVBqv} is about risk, rather than risk-neutral expected returns:
  whenever $\bar{x}_{s}$ and $x^{*}(P_{s})$ differ, \emph{realized} returns of the two strategies
  will differ on any sample path, purely due to differential exposure to risky asset price
  movements.}  In other words, the loss process will have positive quadratic variation along any
given sample path, unless it is essentially equal to the rebalancing strategy.

Since the square root of quadratic variation $X \mapsto \sqrt{[X]_t}$ is a seminorm over the space of
semimartingales $X$, we can view \Cref{cor:lvb} as a ``projection'' result: consider the following
optimization problem over a single sample path
\begin{equation}\label{eq:LVBopt}
  \begin{array}{lll}
    \minimize_{\bar x} & \displaystyle \sqrt{ \left[ \bar{R}(\bar x)_t - V_t \right]_t} \\
    \subjectto & \text{$\bar x$ is an admissible, self-financing trading strategy.}
  \end{array}
\end{equation}
Then, the rebalancing strategy is essentially the unique (up to a set of times
of measure zero) optimal solution to
\eqref{eq:LVBopt}.   In this way,
the rebalancing strategy can be viewed as a projection of the AMM's pool value process onto the
set of self-financing trading strategies that trade at market prices. Moreover, this choice of
benchmark results in a loss process (\LVR) with zero quadratic variation that is
predictable.\footnote{Examining \eqref{eq:LVBeqn}, it is clear that $\LVB(\bar x)$ is a
  super-martingale and that, as per the Doob-Meyer decomposition, it must decompose uniquely into
  an increasing predictable process (the ``compensator'') and a martingale. \LVR is the
  compensator for $\LVB(\bar x)$ independent of the choice of benchmark, while $\Delta(\bar{x})$
  defines the martingale component and depends on the choice of benchmark.}

This derivation thus gives a complementary intuition to the result in \Cref{th:decomposition}. In
equity pricing models, asset alphas are isolated by \emph{projecting} individual asset returns
onto market or factor returns, picking \emph{beta} values which minimize the expected squared
difference between asset and factor returns, and then subtracting the fitted component from
returns. This removes variation due to market or factor risk, leaving a residual that reflects
asset-specific performance.

\Cref{cor:lvb} shows that our approach is analogous: the ``beta-like'' rebalancing strategy is the
quadratic-variation-minimizing projection of the AMM pool value process onto changes in
$P_{s}$. Subtracting the rebalancing strategy's returns yields an ``alpha-like'' residual free of
linear price exposure to the underlying asset.\footnote{We can also draw an analogy to
  options-pricing: the rebalancing strategy corresponds to the \emph{delta} component of LP
  returns. In standard option pricing settings, delta is derived via backward induction and
  depends on unobserved volatility. Computing deltas for AMMs is much simpler: when the price is
  $P_{s}$, the AMM's delta is simply its asset holdings, $x^{*}(P_{s})$.}

The goal of CAPM and FF3 models is not to argue that stocks \emph{should} be held in market-neutral strategies, and the goal of Black-Scholes is not to argue that options \emph{should }be delta-hedged. In both cases, the decompositions serve primarily conceptual purposes. Of course, the decompositions naturally suggest trading strategies: these model-based insights motivate the factor risk management strategies of long-short hedge funds, and the delta-hedging strategies of option market makers. Following this tradition, this paper focuses on the conceptual \pnl decomposition within a frictionless model, leaving a full accounting of the frictional costs and details of implementation to practitioners.


\section{Examples}\label{sec:examples}


\begin{example}[Weighted Geometric Mean Market Maker / Balancer]\label{ex:gmm} Consider the bonding function $f(x,y)\defeq x^{\theta}y^{1-\theta}$, for $\theta\in(0,1)$. Solving the pool value optimization \eqref{eq:pool-min} allows us to obtain the closed-form optimal solutions
\[
x^{*}(P)=L\left(\frac{\theta}{1-\theta}\frac{1}{P}\right)^{1-\theta},\quad y^{*}(P)=L\left(\frac{1-\theta}{\theta}P\right)^{\theta}.
\]
Then,
\[
  V(P)=\frac{L}{\theta^{\theta}(1-\theta)^{1-\theta}}P^{\theta},\quad V''(P)= x^{*\prime}(P)
  =-L\theta^{1-\theta}(1-\theta)^{\theta}\frac{1}{P^{2-\theta}},
\]
and
\[
\ell(\sigma,P)=\frac{\sigma^{2}}{2}\theta(1-\theta)V(P).
\]
\end{example}

The weighted geometric mean market maker generalizes the constant product market maker. For these market makers, the instantaneous \LVR\ normalized per dollar of pool reserves is constant, i.e.,
\begin{equation}
\frac{\ell(\sigma,P)}{V(P)}=\frac{\sigma^{2}}{2}\theta(1-\theta).\label{eq:const}
\end{equation}
In fact, with a minor caveat, weighted geometric market makers are the \emph{only} CFMMs for which this is true. We discuss this in \Cref{sec:wgmm}. Finally, observe that \LVR\ is maximized when $\theta=1/2$, and goes to zero as $\theta\tends\{0,1\}$.\footnote{See also Proposition~1 of \citet{evans2020liquidity}, evaluating a weighted geometric mean market maker over a finite horizon using risk-neutral pricing.}

\begin{example}[Constant Product Market Maker / Uniswap~v2]\label{ex:cpm} Taking $\theta=1/2$ in \Cref{ex:gmm}, we have that
\begin{equation}
V(P)=2L\sqrt{P},\quad\ell(\sigma,P)=\frac{L\sigma^{2}}{4}\sqrt{P},\quad\frac{\ell(\sigma,P)}{V(P)}=\frac{\sigma^{2}}{8}.\label{eq:cpmmexpressions}
\end{equation}
\end{example}

This example shows that the constant product market maker admits particularly simple expressions
for $\LVR$: $\ell(\sigma,P)/V(P)$, the loss per unit time as a fraction of mark-to-market pool value, is
simply $1/8$ times the instantaneous variance. This formula is straightforward to apply
empirically: for example, if the ETH-USDC volatility is $\ensuremath{\sigma=5\% \text{ (daily)}}$,
this formula implies that the ETH-USDC LP pool loses approximately
$\sigma^{2}/8=3.125\text{ (bp)}$ in pool value to $\LVR$ daily.

\begin{example}[Uniswap~v3 Range Order]\label{ex:univ3} Given liquidity parameter $L \geq 0$ and
  prices in the given range $[P_{a},P_{b})$, consider the bonding function of the ``range order''
  \citet{adams2021uniswap},
\[
f(x,y)\defeq\left(x+L/\sqrt{P_{b}}\right)^{1/2}\left(y+L\sqrt{P_{a}}\right)^{1/2}.
\]
Solving the pool value optimization 
 \eqref{eq:pool-min},
\[
x^{*}(P)=L\left(\frac{1}{\sqrt{P}}-\frac{1}{\sqrt{P_{b}}}\right),\quad y^{*}(P)=L\left(\sqrt{P}-\sqrt{P_{a}}\right).
\]
Then, for $P\in(P_{a},P_{b})$,
\[
V(P)=L\left(2\sqrt{P}-P/\sqrt{P_{b}}-\sqrt{P_{a}}\right),
%
\quad V''(P)= x^{*\prime}(P) =-\frac{L}{2P^{3/2}},
\]
so that
\[
\ell(\sigma,P)=\frac{L\sigma^{2}}{4}\sqrt{P}.
\]
\end{example}

Observe that the instantaneous \LVR\ is the same in \Cref{ex:cpm}. However, the pool value $V(P)$ is lower. Indeed $V(P)\tends0$ if $P_{a}\uparrow P$ and $P_{b}\downarrow P$, so
\[
\lim_{\substack{P_{a}\tends P\\
P_{b}\tends P
}
}\frac{\ell(\sigma,P)}{V(P)}=+\infty,
\]
i.e., the instantaneous \LVR\ per dollar of pool reserves can be arbitrarily high in this case, if
the liquidity range is sufficiently narrow. This is consistent with the idea that range orders
``concentrate'' liquidity.

\begin{example}[Uniswap~v3 Pool]\label{ex:univ3-pool}
  A ``concentrated liquidity'' pool such as Uniswap~v3 \citep{adams2021uniswap} aggregates the liquidity across a set of range
  orders $i=1,\ldots,N$, where order $i$ is a range order with liquidity $L_i$ over
  range\footnote{In addition, the endpoints of each price range may be constrained to
    predefined endpoints or ``ticks''.}
  $[P^{(i)}_a,P^{(i)}_b)$. Following \Cref{ex:univ3}, for each order $i$, we have that
  \begin{align}
  x^*_i(P)
  & =
    \begin{cases}
      \displaystyle
      L_i \left( \frac{1}{\sqrt{P^{(i)}_a}} - \frac{1}{\sqrt{P^{(i)}_b}} \right),
      & \text{if $P < P^{(i)}_a$}, \\
      \displaystyle
      L_i \left( \frac{1}{\sqrt{P}} - \frac{1}{\sqrt{P^{(i)}_b}} \right),
      & \text{if $P \in [P^{(i)}_a,P^{(i)}_b)$}, \\
      0, & \text{if $P \geq P^{(i)}_b$}, \\
    \end{cases} \\
  y^*_i(P)
  & =
    \begin{cases}
      0, & \text{if $P < P^{(i)}_a$}, \\
      \displaystyle L_i \left( \sqrt{P} - \sqrt{P^{(i)}_a} \right),
         & \text{if $P \in [P^{(i)}_a,P^{(i)}_b)$}, \\
      \displaystyle L_i \left( \sqrt{P^{(i)}_b} - \sqrt{P^{(i)}_a} \right),
         & \text{if $P \geq P^{(i)}_b$},
  \end{cases}
  \\
  V_i(P) & =
  \begin{cases}
    L_i \left( \frac{1}{\sqrt{P^{(i)}_a}} - \frac{1}{\sqrt{P^{(i)}_b}} \right) P,
    & \text{if $P < P^{(i)}_a$}, \\
    L_i \left( 2\sqrt{P} - P/\sqrt{P^{(i)}_b} - \sqrt{P^{(i)}_a} \right),
    & \text{if $P \in [P^{(i)}_a,P^{(i)}_b)$}, \\
    L_i \left( \sqrt{P^{(i)}_b} - \sqrt{P^{(i)}_a} \right),
    & \text{if $P \geq P^{(i)}_b$},
  \end{cases}
  \\
  \label{eq:vpp}
  V''_i(P) = x_i^{*\prime}(P) & =
  \begin{cases}
    \displaystyle
    -\tfrac{1}{2} L_i P^{-3/2},
    & \text{if $P \in [P^{(i)}_a,P^{(i)}_b)$}, \\
    0, & \text{otherwise.}
  \end{cases}
\end{align}
The aggregate reserves and pool value are given by \citep{jason_exchange_complexity}
\[
  x^*(P) = \sum_{i=1}^N x^*_i(P),\quad
  y^*(P) = \sum_{i=1}^N y^*_i(P),\quad
  V(P) = \sum^N_{i=1} V_i(P).
\]
When we compute (instantaneous) \LVR, we need marginal liquidity $V''(P) = x^{*\prime}(P)$. Observe from
\eqref{eq:vpp} that $V''_i(P)$ is zero when order $i$ is not in-range, i.e., when
$P \notin [P^{(i)}_a,P^{(i)}_b)$. Hence, only in-range orders contribute to \LVR, i.e.,
\[
  \ell(\sigma,P) = -\frac{\sigma^2 P^2}{2} V''(P)
  = -\frac{\sigma^2 P^2}{2} \sum_{i=1} V''_i(P)
  = \frac{\sigma^2}{4} \sqrt{P} \times \underbrace{\sum_{i=1}^N L_i
    \mathbf{1}_{\left\{P\in [P^{(i)}_a,P^{(i)}_b)\right\}}}_{\defeq \bar L(P)}.
\]
\end{example}
The quantity $\bar L(P)$ is the \emph{aggregate in-range liquidity}, given
the current price $P$. From the perspective of the set of range orders which comprises the pool,
this is a sufficient statistic for computing \LVR.

\begin{example}[Linear Market Maker / Limit Order]\label{ex:limit} For $K>0$, consider the linear bonding function $f(x,y)\defeq Kx+y$. Solving the pool value optimization \eqref{eq:pool-min},
\[
x^{*}(P)=\begin{cases}
L/K & \text{if \ensuremath{P<K},}\\
0 & \text{if \ensuremath{P\geq K},}
\end{cases}\quad y^{*}(P)=\begin{cases}
0 & \text{if \ensuremath{P<K},}\\
L & \text{if \ensuremath{P\geq K}.}
\end{cases}
\]
Hence, this pool can be viewed as similar to a resting limit order\footnote{While the linear market maker is \emph{statically} identical to a resting limit order, observe that they are \emph{dynamically} different. In particular, once the price level $K$ is crossed, in a traditional LOB, the limit order is filled and removed from the order book. With a linear market maker, the order remains in the pool at the same price and quantity, but with opposite direction. This merely refers to a superficial change of the \emph{default strategy} after a trade execution.} that is, depending on the relative value of the price $P_{t}$ versus limit price $K$, either an order to buy (if $P_{t}\geq K$) or an order to sell (if $P_{t}<K$) up to $L/K$ units of the risky asset at price $K$. In this case,
\[
V(P)=L\min\left\{ P/K,1\right\} .
\]
Observe that $V(\cdot)$ does not satisfy the smoothness requirement of \Cref{as:smooth-v}
Part~\ref{pt:smooth-v}: the first derivative is discontinuous at the limit price $P=K$. Thus, the
characterization of \Cref{lem:lvr} does not
apply.\footnote{\label{non-locally-smooth-footnote}Note that the pool value function remains
  concave and thus the pool value process is a super-martingale. Hence, from the Doob-Meyer
  decomposition, a non-negative monotonic running cost process (analogous to \LVR) exists. However, this process is not described by \eqref{eq:lvr}--\eqref{eq:ilvr}. Instead, it can be constructed using the concept of ``local time'' and the It\^o-Tanaka-Meyer formula, but we will not pursue such a generalization here \citep[see, e.g.,][]{carr1990stop}.} \end{example}




\section{Empirical Analysis}

\label{sec:empirics}

We now bring our model to data, showing how to separate the ``alpha-like'' and ``beta-like'' components of LP \pnl in practice. This allows us to answer a number of questions. First, how large are the alpha-like and beta-like components, respectively, in terms of their contribution to the variance of returns? Second, how sensitive are the results to the frequency at which we rebalance the AMM's holdings? Third, as we rebalance at increasingly high frequencies, does the result converge to the ``alpha-like'' microstructural component of returns --- fees minus adverse selection costs --- as our model predicts?

\subsection{Methodology}

\begin{figure}
  \begin{center}
    \includegraphics[width=\textwidth]{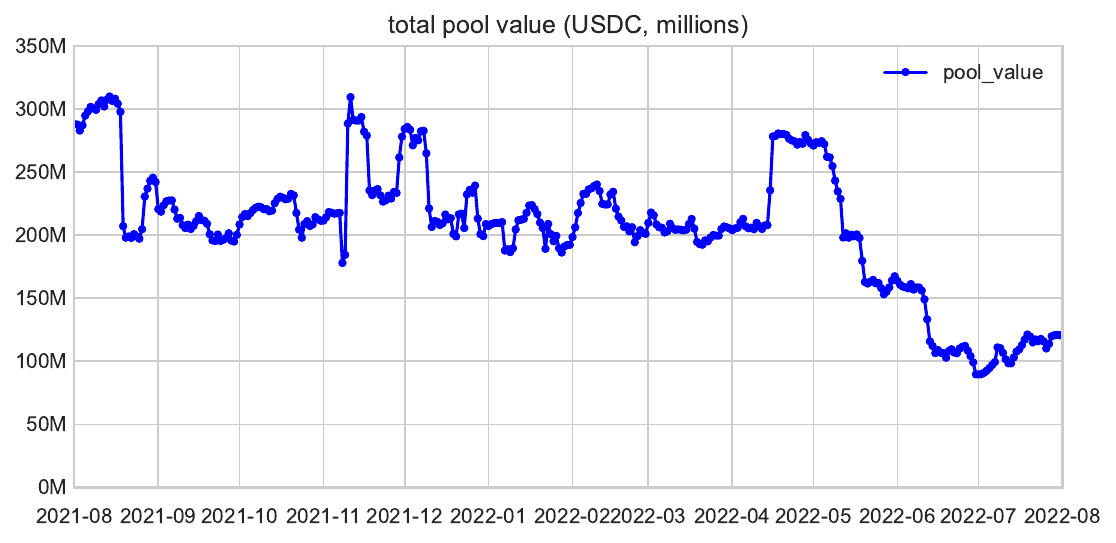}
  \end{center}
  \caption{The daily average pool value of the Uniswap~v2 WETH-USDC pool. \label{fig:pool_value}}
\end{figure}

We bring the model to data using the WETH-USDC trading pair\footnote{``WETH'', or ``wrapped ETH''
  is a variation of ETH that is compliant with the ERC-20 token standard. Conversion between the
  two is a straightforward operation with a minimal fixed cost. For our purposes, we will view ETH
  and WETH as equivalent.} on Uniswap~v2 for the period from August 1, 2021 to July 31, 2022. The
value of the reserves of this pool over the period are shown in \Cref{fig:pool_value}.
Details of the data sources we use, and how we measure various quantities, are described in \Cref{sec:measurement}.

Repeating \eqref{eq:PnLdecomp}, we have
\begin{equation}
  \overbrace{\text{LP \pnl}_{t}-\underbrace{\int_{0}^{t}x^{*}(P_{s})\,dP_{s}}_{\textrm{market\
        risk}}}^{\text{hedged \pnl}}
  =\underbrace{\FEE_{t}-\LVR_{t}}_{\text{fees minus \LVR}}.\label{eq:empiricaldecomp}
\end{equation}
The left side of \eqref{eq:empiricaldecomp} can be thought of as the $\pnl$ from a delta-hedged LP position: the
$\pnl$ of the LP position, minus that of the rebalancing strategy. This quantity can be estimated
empirically under very weak assumptions:
\begin{itemize}
  \item We can approximate the integral in  \eqref{eq:empiricaldecomp} using \emph{discrete rebalancing strategies}; for a fixed interval length $\Delta$, we define:
\begin{equation}
\text{RB\ \ensuremath{\pnl_{t}}}\defeq\sum_{s=0,\Delta,2\Delta,\ldots}^{t}\Delta\text{RB\ \ensuremath{\pnl_{t}}},\label{eq:empiricsRBPnL}
\end{equation}
\begin{equation}
\Delta\text{RB\ \ensuremath{\pnl_{t}}}\defeq x^{RB}_t \left(P_{t+\Delta t}-P_{t}\right).\label{eq:empiricsDeltaRBPnL}
\end{equation}
Here, $x^{RB}_t$ is the quantity of the risky asset in the pool reserves at time $t$.
Mathematically, \eqref{eq:empiricsRBPnL} is a Riemann sum approximation for rebalancing strategy
profits. Economically, \eqref{eq:empiricsRBPnL} can be thought of as the payoffs of a strategy
which delta-hedges at market prices to match the AMM's holdings at time intervals $\Delta$. This
strategy holds $x^{RB}_t$ on the interval $[t,t+\Delta t)$, and thus makes dollar profits
equal to the holdings times the change in prices $P_{t+\Delta t}-P_{t}$ on this interval.
\item We measure $\text{LP \pnl}_{t}$, as our model suggests, essentially as $P_{t}x_{t}+y_{t}$,
  using the observed pool holdings $x_{t},y_{t}$ and the CEX price $P_{t}$. In practice, pool
  reserves adjust over time through ``mints'' and ``burns'' of LP positions; we value these mints
  and burns periodically to CEX prices. Details of these calculations are provided in
  \Cref{sec:measurement}.
\end{itemize}

\subsection{Empirical Results}

\begin{figure}
  \begin{subfigure}[t]{\textwidth}
    \centering
    \includegraphics[width=0.9\textwidth]{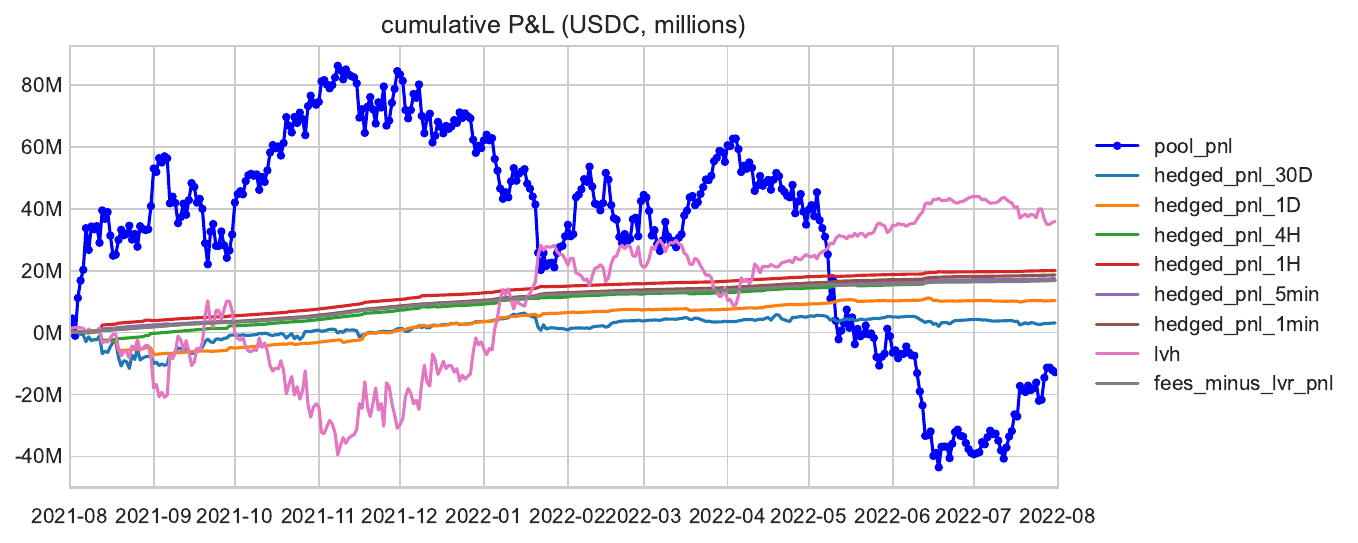}
    \caption{Including the aggregate pool \pnl series.\label{fig:pool_plus_hedged_pnl}}
  \end{subfigure}

  \bigskip

  \begin{subfigure}[t]{\textwidth}
    \centering
    \includegraphics[width=0.9\textwidth]{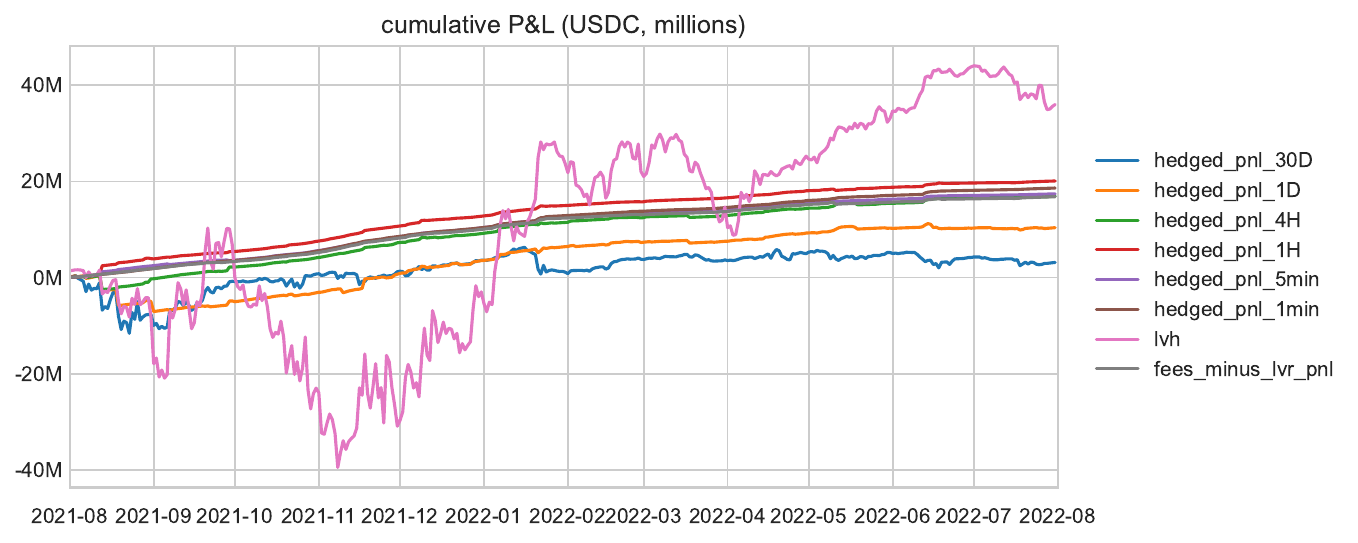}
    \caption{Excluding the aggregate pool \pnl series.\label{fig:hedged_pnl}}
  \end{subfigure}
  \caption{Cumulative pool \pnl, delta-hedged \pnl, and predicted \pnl from our expressions for
    $\LVR$, for the Uniswap~v2 WETH-USDC trading pair. In the first panel, the \texttt{pool\_pnl}
    series shows the raw \pnl of the LP position.  In both panels, the various \texttt{hedged\_pnl}
    series show delta-hedged LP \pnl, that is, the \pnl from a long position in the pool, and a
    short position in the rebalancing strategy, rebalanced at various frequencies (daily, every
    four hours, every hour, every five minutes).  The \texttt{fees\_minus\_lvr\_pnl} series shows fees
    minus $\LVR$, this is the delta-hedged \pnl predicted by our theory.
    The data source
    for prices is Binance, and LP \pnl is calculated using data on Uniswap trades, mints, and
    burns from the Ethereum blockchain. Details of how we calculate these quantities are in
    \Cref{sec:measurement}.\label{fig:pnl}}
\end{figure}

\Cref{fig:pnl} shows the results of our analysis. In \Cref{fig:pool_plus_hedged_pnl}, the
\texttt{pool\_pnl} series shows unhedged LP pool \pnl. The pool's value is fairly volatile: daily
gains or losses of over \$10 million USD (on a pool reserves whose value varies between
$\$100$--$\$300$ million USD, cf.\ \Cref{fig:pool_value}) are common. However, the vast majority
of this risk is simply market risk: the pool holds a long position in ETH, whose price is very
volatile over the period we study (cf.\ \Cref{fig:realized_volatility}). The
\texttt{\texttt{hedged\_pnl}} series show the hedged \pnl from \eqref{eq:empiricaldecomp},
rebalancing over various frequencies. Visually, these series are so much smaller than the aggregate
\texttt{pool\_pnl} series that their fluctuations are almost invisible in
\Cref{fig:pool_plus_hedged_pnl}.

\begin{table}
  \centering
  \begin{tabular}{llrrr}
  \toprule
  \multicolumn{1}{c}{ \textbf{\sffamily series} }
  & \multicolumn{1}{c}{ \textbf{\sffamily rebalancing freq} }
  & \multicolumn{1}{c}{ \textbf{\sffamily avg return} }
  & \multicolumn{1}{c}{ \textbf{\sffamily stdev return} }
  & \multicolumn{1}{c}{ \textbf{\sffamily Sharpe ratio} } \\
  &
  & \multicolumn{1}{c}{ (\%, annual) }
  & \multicolumn{1}{c}{ (\%, annual) }
  & \multicolumn{1}{c}{ (annual) } \\
  \midrule
  \texttt{pool\_pnl}             & unhedged & -6.20 & 42.08 & -0.15 \\
  \texttt{lvh}                   & once &   17.46 &  27.08 &  0.64 \\
  \texttt{hedged\_pnl\_30D}      & 30 day &  1.53 &  7.27 &  0.21 \\
  \texttt{hedged\_pnl\_1D}       & 1 day & 5.04 &  2.87 &  1.75 \\
  \texttt{hedged\_pnl\_4H}       & 4 hour & 8.21 &  1.50 &  5.47 \\
  \texttt{hedged\_pnl\_1H}       & 1 hour & 9.75 &  0.90 & 10.81 \\
  \texttt{hedged\_pnl\_5min}     & 5 min & 8.45 &  0.47 & 18.16 \\
  \texttt{hedged\_pnl\_1min}     & 1 min & 9.04 &  0.39 & 23.33 \\
  \texttt{fees\_minus\_lvr\_pnl} & continuous limit & 8.16 &  0.48 & 17.03 \\
  \bottomrule
\end{tabular}


  \caption{Overall return statistics for the cumulative \pnl series of
    \Cref{fig:pnl}.\label{tab:ret} This table shows the standard deviation of hedged \pnl, where hedging is done
    at interval lengths ranging from every minute, to every month. The \texttt{lvh} series
    considers hedging a single time, using ETH holdings at the start of our sample period, on
    August 1 2021, corresponding to ``impermanent loss'' using a single starting
    point. `The \texttt{pool\_pnl} series corresponds to raw LP \pnl. \label{tab:rebalfreq}}
\end{table}

\Cref{fig:hedged_pnl} zooms in, focusing on the hedged $\pnl$.  All of these series are
substantially less volatile than the unhedged \pnl. This is quantified in the return statistics in
\Cref{tab:rebalfreq}. Quantitatively, the standard deviation of returns of 1-minute hedged \pnl is
less than 1\% of the standard deviation of returns of the unhedged \pnl; in variance terms,
0.009\% of the variation in returns in unhedged \pnl comes from fees and adverse selection, and
the remaining 99.991\% simply reflects ETH price movements.

As the rebalancing intervals, hedged LP \pnl becomes increasingly more volatile, reflecting market
risk accrued within intervals. This is consistent with \Cref{cor:lvb}, stating that the
rebalancing strategy corresponds to the variance-minimizing projection of LP \pnl on market risk:
trading strategies which are increasingly coarse approximations to the rebalancing strategy thus
leave increasing amounts of variance within the hedged LP time series.

In the \texttt{lvh} series of \Cref{fig:pnl} and \Cref{tab:rebalfreq}, we consider a benchmark which
holds ETH holdings at the start of the sample and \emph{never} rebalances, corresponding to
``impermanent loss'' (cf.~\Cref{sec:lvr-lvh}). The volatility of ``impermanent loss'' measure is
of the same order of magnitude at the unhedged pool. This illustrates why ``impermanent loss''
fails to account for market risk exposures: it perfectly hedges market risk at the start of the
time interval, but accrues market risk as prices and holdings move over time; the result is a
series that is almost as polluted by market risk as raw LP \pnl.

\subsection{Testing the Theory}\label{sec:testtheory}

\begin{figure}
  \begin{center}
    \includegraphics[width=\textwidth]{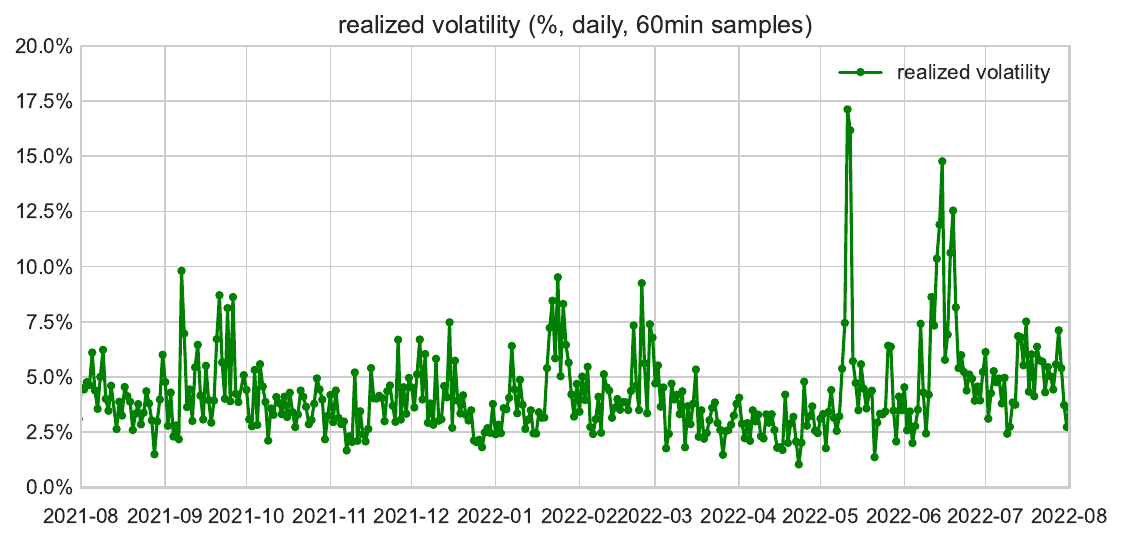}
  \end{center}
  \caption{Daily realized volatility for the ETH-USDC pair, computed from Binance minutely closing
    prices, sampled at 60 minute intervals.
    \label{fig:realized_volatility}}
\end{figure}

Our model implies that, as we approximate the profits of the rebalancing strategy increasingly
well, the hedged \pnl should converge not to zero, but to the ``alpha-like'', microstructural
component of LP returns on the right side of \eqref{eq:empiricaldecomp}: accrued fees minus
accrued adverse-selection costs. We can test this prediction by directly measuring this series,
and comparing it to high-frequency hedged LP returns. We directly observe $\FEE_{t}$, fees paid
into the LP pool over any given time period.\footnote{Uniswap v2 also allows for assets to be
  borrowed for ``flash loans'' for a fee; however, in v2, these loans are accounted for as swaps,
  thus $\FEE_{t}$ contains revenue from flash loans. We discuss this briefly in
  \Cref{subsec:data}.} For $\LVR_{t}$, since Uniswap v2 is a constant-product CFMM, adverse
selection losses have the particularly simple expression in \eqref{eq:cpmmexpressions} of
\Cref{ex:cpm},
\begin{equation}
\LVR_{t}=\int_{0}^{t}\frac{\sigma_{s}^{2}}{8}\times V(P_{s})\,ds.\label{eq:lvr-cpm}
\end{equation}
We estimate daily realized volatility using Binance prices, see \Cref{fig:realized_volatility}.
We can then compute \LVR simply by plugging in realized daily volatility and pool value to a
version of equation \eqref{eq:lvr-cpm} that is discretized at a daily frequency.

Note that, empirically, we measure the total fees paid by all kinds of traders. This differs slightly from our model, where we assume arbitrage traders pay no fees. Practically, since fees are simply an increasing process which potentially compensates for $\LVR$, whether fees arise from noise trade or arbitrage trade does not substantially impact the returns on LP positions. If we assumed arbitrage traders paid trading fees in the model, this would decrease the amount of arbitrage: instead of prices on the CFMM moving immediately to match CEX prices at all times, prices would have to move more than fees in order for arbitrage trade to have nonnegative payoffs. The analysis of arbitrage profits in the presence of fees is the subject of follow-on work \citep{2023-02-lvr-fee-model}. 

\begin{figure}
  \begin{center}
    \includegraphics[width=\textwidth]{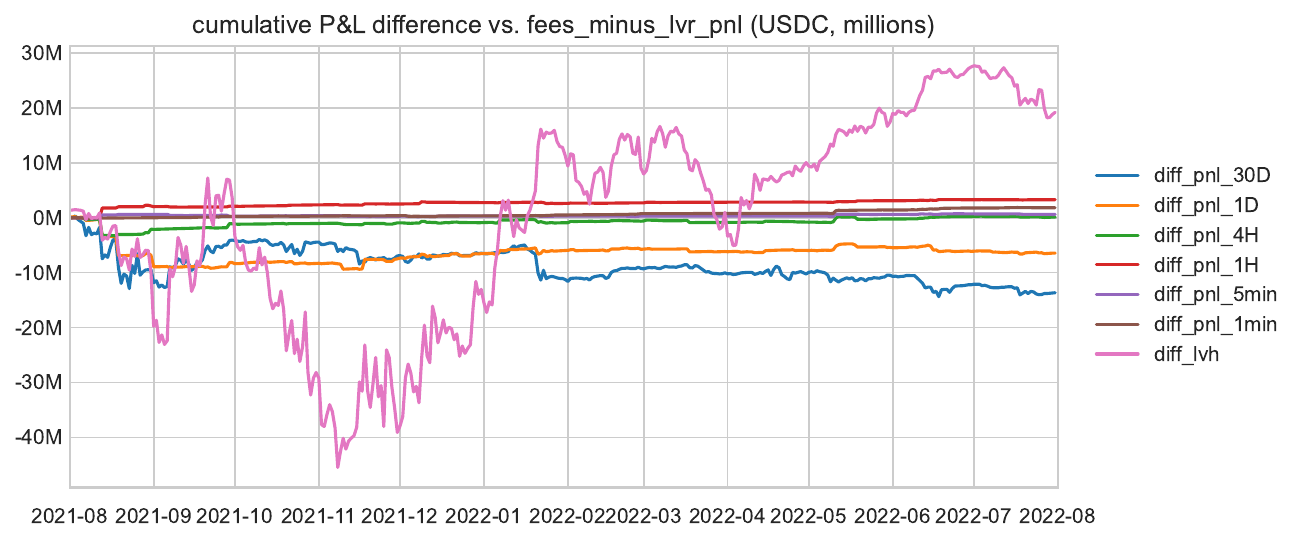}
  \end{center}
  \caption{Cumulative differences between the delta-hedged \pnl (various \texttt{hedged\_pnl}
    series of \Cref{fig:pnl}) and predicted \pnl from our model (\texttt{fees\_minus\_lvr\_pnl}
    series of \Cref{fig:pnl}). \label{fig:diff_pnl}}
\end{figure}

We plot fees minus \LVR as the \texttt{fees\_minus\_lvr\_pnl} series in \Cref{fig:pnl}. Fees minus
\LVR in our model closely track the pattern of hedged LP \pnl. We plot the difference between
these series in \Cref{fig:diff_pnl}. Observe that the gap is smaller for more frequent rebalancing
frequencies. In this way, fees minus \LVR is an approximate continuous time limit of hedged \pnl,
just as our model predicts.

This increases our confidence that
subtracting the ``beta-like'' rebalancing strategy from LP \pnl indeed leaves an ``alpha-like''
residual which reflects the contribution of microstructural forces to AMM LP returns. In practice,
we recommend that empirical researchers primarily construct hedged LP \pnl simply by subtracting
the rebalancing strategy's profits from observed LP \pnl: directly subtracting the payoffs of the
hedging strategy is more straightforward to interpret, and imposes less theoretical structure on
the underlying data.

\subsection{Implications for Empirical Analysis of AMM Returns}

\Cref{fig:pnl} and \Cref{tab:rebalfreq} show that LP \pnl consists almost overwhelmingly of the
``beta-like'' market risk component. This has important implications for empirical analyses of LP
returns: if market risk is not explicitly removed from LP returns, any results will predominantly
reflect market risk rather than microstructural effects.

Suppose a researcher wishes to study the determinants of LP \pnl, and attempts to do so by
regressing unhedged LP returns (or fixed-start-point ``impermanent loss'', which behaves similarly)
on some $x$-variable. The best-case scenario for the researcher is that market risk is pure noise,
uncorrelated with $x$-variables. OLS coefficient standard errors are proportional to the standard
deviation of the regression residual; thus, even in this best case, using unhedged LP returns as
the y-variable leads to standard errors roughly 108 times larger than if hedged returns were
used. Since standard errors scale with the inverse square root of sample size, achieving equal
coefficient standard errors using unhedged LP returns as a $y$-variable would require around
$108^2=11{,}664$ times more data.

In reality, a more severe concern is that market risk may be correlated with any $x$-variables
used. The large relative variance of market risk implies that even a weak correlation between the
$x$-variable and market risk can induce substantial omitted-variable bias, easily swamping any
genuine relationship between $x$ and hedged LP returns.

We therefore recommend that empirical researchers analyzing AMM LP returns implement our results
by using hedged LP returns as regression inputs. Our procedure has very low data requirements: to
construct profits of the rebalancing strategy, \eqref{eq:empiricsDeltaRBPnL}, the researcher only
needs $x_{t}^{RB}$, the time series of risky asset holdings of the CFMM LP, and $P_{t}$, the CEX
price of the risky asset. Our results also appear robust to the choice of rebalancing
frequency. Failing to hedge can have severe consequences: in empirical designs using unhedged LP
returns, any microstructural ``signal'' --- in our model, fees minus adverse selection costs ---
is likely to be overwhelmed by market noise.

A trader attempting to delta-hedge LP returns in practice would of course face various
implementation costs, such as spreads and fees for rebalancing trades, or margin costs for
maintaining a short position in the risky asset. Our goal here is not to precisely estimate the
\emph{practical} profits from hedged LP-ing; rather, it is mainly to \emph{conceptually} decompose
profits into market risk and residual components, analogous to alpha-beta decompositions in
asset-pricing factor models. Following practice in this literature, we thus calculate
$\text{RB\ \ensuremath{\pnl_{t}}}$ assuming frictionless rebalancing.\footnote{While frictions of
  course play a role in practice, large trading firms have a variety of ways of optimizing these
  frictional costs; for example, they may rebalance at lower frequencies, optimize rebalancing
  execution, or ``cross-hedge'' by netting delta exposures across different parts of the
  portfolio, to limit trade and shorting costs. These practical factors imply that any attempt we
  make to quantify the frictional costs of rebalancing is unlikely to be generally accurate across
  all applied settings. }

\subsection{Markouts}

Industry participants frequently use ``markouts'' as an approach to evaluate CFMM LP $\pnl$
\citep[see, e.g.,][]{crocswap_usage_2022}. Markouts, known as ``realized spread'' in the
microstructure literature \citep{huang1996dealer}, essentially attribute profits to trades by
comparing the price of each trade to a future price, usually at some fixed time offset from the
trade (e.g., 10~minutes) either from a CEX, or the CFMM itself. However, delta-hedging and the
markout approach are in a sense closely related: the $\pnl$ of delta-hedged LPing, when rebalanced
at discrete periods, turns out to be exactly equivalent to markout profits, marked to CEX prices
at the end of discrete periods of the same frequency. In other words, the main difference between
delta-hedging and markout analysis is whether the marking price is obtained in a fixed offset of
time in the future after a trade, or based on the ending price of the interval the trade was
contained in. These two are likely to be very close, especially when the markout offset or
delta-hedging interval are short. Thus, delta-hedged LPing can be thought of as a microfoundation
for markout-style analysis.


\section{``Impermanent Loss''}\label{sec:lvr-lvh}

It is common in industry practice to evaluate LP \pnl against simply holding the LP's asset mix at
some fixed start point; this is the loss metric we call ``loss versus holding'' (\LVH) in earlier
sections, and is sometimes called ``impermanent loss'' in industry practice. In the notation of
\Cref{subsec:conttimeprojection}, defining
the benchmark strategy $x_{t}^{\HODL}\defeq x^{*}(P_{0})$, then
$\LVH_t \defeq \LVB_{t}(x_{t}^{\HODL})$ is defined as the loss process relative to this benchmark.
Like \LVR, \LVH has the property that it is non-negative, and can thus also be viewed as a ``cost'':
\begin{proposition}
  For all $t\ge 0$, $\LVH_t \ge 0$.
\end{proposition}
\begin{proof}
  \[
    \LVH_t
    =
    \bar{R}_{t}(x^{\HODL}) - V_t
    =
    \left( P_t x^{*}(P_{0}) + y^{*}(P_{0}) \right)
    -
    \left( P_t x^{*}(P_{t}) + y^{*}(P_{t}) \right),
  \]
  and, by the optimality of \Cref{eq:pool-min} for $P=P_t$, this quantity is always non-negative.
\end{proof}

\LVH seems like a simpler metric than \LVR, since it benchmarks profits to a single start point,
and does not require calculating the profits of a continuously rebalanced benchmark
portfolio. However, this seeming simplicity leads to severe problems with using \LVH in practice.

\subsection{Problems with Loss-Versus-Holding}

\label{subsec:lvhproblems}

First, \LVH fails a basic ``cumulation'' property that is satisfied by most return-like metrics used in finance. Let $\LVH_{t,\tilde{t}}$ denote loss-versus-holding from $t$ to $\tilde{t}$, evaluated against buying and holding the time-$t$ portfolio. We have, in general:
\begin{equation}
\LVH_{t_{1},t_{2}}+\LVH_{t_{2},t_{3}}\neq \LVH_{t_{1},t_{3}}\label{eq:LVHnonadditivity}
\end{equation}
This non-additivity arises because $\LVH_{t_{2},t_{3}}$ is benchmarked against time-$t_{2}$ holdings, instead of time-$t_{1}$ holdings. A simple example appears in the ``reconvergence'' paths $P^{UD}$ and $P^{DU}$ on \Cref{fig:binomtree}. The AMM makes trades and incurs positive \LVH on each individual time step: $\LVH_{0,1}>0$ and $\LVH_{1,2}>0$. However, the $t=2$ portfolio is equal to the $t=0$ portfolio so we have $\LVH_{0,2}=0$! The econometrician faces a paradox: the AMM appears to suffer positive losses at each individual time step, but no loss over the interval as a whole. Relatedly, \LVH is not monotonic: adverse selection costs may appear to reverse as prices reconverge, even though the AMM has incurred genuine adverse selection costs at every step. By contrast, \LVR is monotonically non-decreasing, reflecting the economically natural property that adverse selection costs accumulate irreversibly over time.

Second, the sign of the change in $\LVH$ can vary, depending on the start point chosen. Consider an AMM with a fixed level of liquidity; suppose the AMM holds $x=100$ of the risky asset at $t_{1}$, and $x=200$ at $t_{2}$. Consider a future time period $\left[t_{3},t_{4}\right]$ during which holdings increase from 150 to 160. Then:
\[
\LVH_{t_{1},t_{4}}-\LVH_{t_{1},t_{3}}<0
\]
\[
\LVH_{t_{2},t_{4}}-\LVH_{t_{2},t_{3}}>0
\]
That is, $\LVH$ seems to \emph{increase} relative to the $t_{1}$ benchmark, but \emph{decrease} relative to the $t_{2}$ benchmark! Both interpretations are internally consistent; however, the econometrician cannot unambiguously conclude whether, in the time period $\left[t_{3},t_{4}\right]$, the AMM made or lost money. The magnitude, and even the sign, of the incremental profit depends on an arbitrarily chosen reference point.

Third, \LVH is path-independent: for any CFMM, $\LVH_{0,T}$ depends only on the initial and final prices $P_0$ and $P_T$, not on the path of prices between them. This is economically problematic, since adverse selection costs should naturally depend on the price path. An LP should lose more to informed trading during a period of high volatility than during a period of low volatility, even if prices start and end at the same levels. By contrast, \LVR is explicitly path-dependent: for a constant-product AMM, \eqref{eq:lvr-cpm} shows that \LVR accumulates faster during high-volatility periods, consistent with the intuition that adverse selection is more costly when prices are more volatile.

Fourth, \LVH conflates the ``alpha-like'' microstructural component of LP returns --- adverse
selection costs specific to the AMM's market-making function --- with the ``beta-like'' market
risk arising from the AMM's asset holdings.
Because \LVH benchmarks against fixed holdings, it accrues market risk as the AMM's holdings
diverge from the benchmark over time (cf.~\Cref{cor:lvb}). As our empirical analysis in \Cref{sec:empirics} shows, this beta-like component is orders of magnitude larger than the alpha-like component in variance terms, and can easily overwhelm any microstructural signal. \LVR, by contrast, cleanly separates these two forces: the rebalancing strategy absorbs all beta-like market risk, isolating the alpha-like microstructural component of LP returns.

A natural way to fix these problems is to re-benchmark performance in each short time interval against holdings at the start of that interval. Fixing some period length $\Delta$, we could define a loss metric $L$, as: 
\begin{equation}
L_{t_{1},t_{2}}\defeq \LVH_{t_{1},t_{1}+\Delta}+\LVH_{t_{1}+\Delta,t_{1}+2\Delta}+\ldots+\LVH_{t_{2}-\Delta,t_{2}}\label{eq:LVHdiscrete}
\end{equation}
That is, over each length-$\Delta$ interval, AMM payoffs are compared to holding the portfolio fixed at the beginning of that interval. This construction guarantees additivity:
\begin{equation}
L_{t_{1},t_{2}}+L_{t_{2},t_{3}}=L_{t_{1},t_{3}}\label{eq:LVRadditivity}
\end{equation}
But \eqref{eq:LVHdiscrete} is exactly the definition of \LVR! Thus, $\LVR$ can be thought of as simply a \emph{time-discretized, path-consistent} version of \LVH: it simply updates the benchmark portfolio at every interval as prices and holdings evolve.

\subsection{``Impermanent Loss'' in the Empirical AMM Literature}

\label{subsec:empiricallit}

In the one-period models used in the majority of the theoretical literature on AMMs, ``impermanent loss'' has a simple and unambiguous interpretation. However, empiricists who wish to measure ``impermanent loss'' in practice must make an important measurement choice: whether to treat time as one long period, or several shorter periods.

In the absence of theoretical guidance, both choices have been made in the empirical literature. \citet{augustin2024reaching} appears to simply fix a start point and calculate impermanent loss based on holdings at that point in time.\footnote{\citet{augustin2024reaching} write: ``The aggregate h-period return to yield farming is thus composed of four components associated with capital gains, impermanent losses, trading fees, and realized farm yields...'' The following expression (8) then contains a difference of return metrics $R_{t,t+h}^{A}$ and $R_{t,t+h}^{B}$, leading us to believe that the holdings are fixed at a start point $t$, which appears to be set to September 23, 2020 for all pools.} Other papers, such as \citet{fang2024learning} and \citet{aquilina2024decentralised}, appear to calculate impermanent loss on a number of shorter time intervals, and add the resulting losses up; specifically, %
we believe \citet{aquilina2024decentralised} calculates impermanent loss using daily-updated holdings benchmarks,\footnote{On page 27, \citeauthor{aquilina2024decentralised} write: ``We assess Fee Yield and Impermanent Loss on a daily basis, i.e. an LP can decide whether or not to provide liquidity at the end of each day''.} and that the main specifications in \citet{fang2024learning} correspond to weekly-rebalanced impermanent loss.\footnote{\citet{fang2024learning} writes: ``For the pool-level analyses, I aggregate transactions for each pool at daily, weekly, or monthly frequency. I focus on the weekly frequency because this appears to be the modal horizon for liquidity adjustment.'' We note that \eqref{eq:LVHnonadditivity} implies that there is no guarantee that daily-frequency losses thus calculated add up to weekly or monthly frequency results. \citeauthor{fang2024learning} also writes: ``Instead of loss-vs-holding, Milionis et al. (2022) show that LP returns can be characterized by loss-vsrebalancing (\LVR). My main results are qualitatively the same when using \LVR.'' However, results using \LVR are not reported in the paper. This statement is plausible, since according to our definitions, the weekly-updated-holdings ``impermanent loss'' measure used in \citet{fang2024learning} calls is exactly equivalent to what we call ``loss-versus-rebalancing'', rebalanced at a weekly frequency.} Some papers, such as \citet{heimbach2022risks}, do not specify whether the holdings benchmark is updated periodically, or what the start point is. 

All these papers refer to the object they measure as ``impermanent loss'', implicitly assuming that the underlying economic object measured is insensitive to these measurement decisions. This is incorrect. ``Impermanent loss'' measured using periodically updated holdings, versus fixed holdings, are substantively different metrics. The periodically-updated measure converges to fees minus adverse selection costs, for reasonably high update frequencies. The fixed-start-point measure, in our analysis in \Cref{sec:empirics}, has over a hundred times greater variance than the periodically-updated measure, and mainly reflects movements in ETH prices rather than any microstructural forces. The periodically-updated measure does not add up to the fixed-start measure, and the latter can easily be zero while the former is positive. 

In our view, there is a missing link between the theoretical and empirical literature on AMMs: \emph{market risks} are not salient in the one-period models used in most of the theoretical literature. As we showed in \Cref{fig:binomtree}, ``impermanent loss'' (loss versus holdings) and \LVR are identical over a single time step; the difference only emerges at the second step, where the AMM's holdings deviate from the fixed-holding benchmark, and thus \LVH begins to accrue market risk. This point has not been explicitly made in the theoretical AMM literature, allowing market risk to creep unnoticed into the metrics used in the empirical AMM literature. Our paper aims to fill this missing link: a simple \emph{dynamic} model of AMMs allows us to clearly separate the distinct contributions of microstructural forces and market risk to AMM \pnl.

Our paper gives clear guidance on how the empirical literature should proceed. If the goal of analysis is to measure \emph{microstructural effects} --- the performance of the AMM in its market-making capacity --- the market-risk\emph{ }component of returns should be treated as a nuisance parameter to be eliminated, akin to how factor models are used to remove known sources of predictability from returns in equity-pricing settings. This can be done simply by subtracting the profits of the rebalancing strategy from pool \pnl, as we do in \Cref{sec:empirics}. This is approximately the approach taken in \citet{fang2024learning} and \citet{aquilina2024decentralised}, though --- aggressively extrapolating from our empirical analysis --- both papers could benefit from slightly more frequent rebalancing, as we find that market risk still dominates microstructural forces at daily and weekly rebalancing frequencies. The loss measure obtained through this approach is mathematically and economically distinct from ``impermanent loss'' based on a fixed start point. We suggest, of course, calling this updated-holdings measure ``loss versus rebalancing'' instead of ``impermanent loss''. 

Of course, different measures serve different purposes, and we cannot say in any formal sense that \LVR is the universally ``best'' metric. However, our model does suggest that fixed-start-point ``impermanent loss'' is simply equal to \LVR, plus the market risk exposures of a well-defined trading strategy, and the market risk component may be much larger than microstructural effects in variance terms. Thus, we believe papers which use fixed-start ``impermanent loss'' should make an argument for why it is appropriate to include this market risk, for the particular question they consider. As a minimum standard, papers should carefully specify the start-point and updating-frequency decisions they make in their implementation of metrics labelled ``impermanent loss''.


\section{Option Pricing}

\label{sec:options}

We have shown that CFMM LPs behave like a bet on volatility, in the sense that $\LVR$ is large
when volatility is high. In this section, we briefly discuss the relationship of CFMM LPs to three
classical and inter-related ways that volatility can be traded \citep{carr2001towards}: static
(European) options positions, dynamic trading strategies, and variance swaps. In
\Cref{subsec:binomtree}, we also demonstrate the equivalences between CFMM LP positions and these
concepts in a simple two-step binomial tree model.

\subsection{Static Options Positions}

Our results are related to \citet{clark2020replicating}, \citet{fukasawa2022weighted}, and
\citet{deng2023static}, who show that, over any finite time horizon, an AMM LP position's payoff,
ignoring fees, can be replicated by shorting a bundle of European options. Technically, this
follows from the facts that the CFMM's asset position and value are both path-independent: if the
price is $P_{T}$ at time $T$, the CFMM holds $x^{*}(P_{T})$ of the risky asset and has pool value
$V_{T}=V(P_{T})$ corresponding to the final ``payoff function'' $V(\cdot)$, regardless of the path that prices took to reach $P_{T}$. More intuitively, at
any time $T$, the CFMM simply offers a menu of quantities of the asset to buy or sell at any given
price, identically to a portfolio of European options. Ignoring fees, the CFMM exactly breaks even
if prices do not move, $P_{T}=P_{0}$, and loses money otherwise; hence, the CFMM LP position is
essentially equivalent to \emph{giving away} a bundle of European options. This intuition is
consistent with the fact that the $V(\cdot)$ is a \emph{concave} function (cf.\ \Cref{le:envelope}).

Expected $\LVR$ until period $T$ can be thought of as the value of the European options given away. This analogy gives another intuition for the comparative statics of expected $\LVR$. European options are worth more when volatility is higher, so $\LVR$ is increasing in the volatility of the underlying asset. When the marginal liquidity of the AMM bonding curve is greater, the replicating portfolio of European options is larger: AMMs that trade more aggressively essentially give away more European options, also increasing $\LVR$.

\textbf{}%
As previous papers have discussed, the European option replication result also implies that, over any finite time horizon, the exposure of the AMM LP position to underlying prices can be \emph{totally} hedged, by taking a long position in the replicating bundle of European options. This trade --- a long position in the AMM LP, plus a short position in the replicating bundle of European options --- is essentially a \emph{trading fee swap}, betting on whether accrued trading fees from time $0$ to $T$ are greater than European option premia of the replicating portfolio at time $0$. The trader enters an LP position, and pays a premium for buying the replicating bundle of European options upfront. The AMM LP position then loses no money from price movements; the total position profits if the accrued trading fees until time $T$ are greater than the European option premia paid upfront, and loses otherwise.

\subsection{Dynamic Trading Strategies}

Classic options theory implies that static option positions are equivalent to dynamically trading
the underlying asset in a certain way. The static option position is a combination of short
straddles and strangles, selling out-of-money calls and puts. This position is equivalent to a
dynamic trading strategy which sells the asset when prices increase, and buys when prices
decrease. This is exactly what the AMM LP position does: observe that, from \Cref{le:envelope} Part~\ref{pt:Vpp}, $x^{*}(\cdot)$ is non-increasing. If prices decrease slightly from $P_{0}$ to $P_{t}<P_{0}$, the rebalancing strategy responds by buying the risky asset. The rebalancing strategy thus makes a profit, relative to simply holding the initial position $x^{*}\left(P_{0}\right)$, if prices increase back to $P_{0}$, and makes a loss if prices decrease further from $P_{t}$. This argument holds symmetrically for price decreases, implying that the rebalancing strategy makes losses if prices diverge from $P_{0}$, and profits when prices make small movements away from $P_{0}$ and back. In the special case where the risky asset's price is a random walk, the rebalancing strategy thus breaks even on average. In contrast, when prices move away from $P_{0}$ and back, the CFMM reverts to the initial value $V\left(P_{0}\right)$, exactly breaking even: there is no profit from price convergence, to offset the losses the CFMM makes when prices diverge from $P_{0}$.

\subsection{Variance Swaps}

Finally, as discussed by \citet{fukasawa2022weighted}, variance can be traded directly by trading
swaps on realized variance. The VIX is such a contract, operating on a fixed finite time
horizon. Applying \Cref{le:envelope} Part~\ref{pt:Vpp}, the instantaneous \LVR of \eqref{eq:ilvr} can be re-written as 
\[
\ell(\sigma,P)=\tfrac{1}{2}\times(\sigma P)^{2}\times|x^{*\prime}(P)|.
\]
Here, the first component, $(\sigma P)^{2}$, is the instantaneous variance or quadratic variation
of the price, i.e., for small $\Delta t$,
$\Var[P_{t+\Delta t}|P_{t}=P]\approx(\sigma P)^{2}\,\Delta t$. Recalling that $x^{*}(P)$ is the
total quantity of risky asset held by the pool if the price is $P$, the second component,
$|x^{*\prime}(P)|$ corresponds to the \emph{marginal} liquidity available from the pool at price
level $P$. Now, integrating over time, we have that
\[
  \LVR_{t}=\tfrac{1}{2}\int_{0}^{t}(\sigma_s P_{s})^{2}\times|x^{*\prime}(P_{s})|\,ds
= \tfrac{1}{2}\int_{0}^{t}|x^{*\prime}(P_{s})|\, d[P]_s
  ,\quad\forall\ t\geq0.
\]
This expression is the payoff of the floating leg of a continuously sampled generalized variance swap \citep[see, e.g.,][]{carr2009volatility}, specifically a price variance swap that is weighted by marginal liquidity.


\section{Discussion and Implications}

\label{sec:discussion}

Besides their positive value for understanding and quantifying the losses from AMM LPing, our results also suggest ways that these losses could be reduced or eliminated. We showed that CFMM LPs lose money from price slippage: when CEX prices move, CFMM quotes become ``stale'' and are vulnerable to sniping by rebalancing arbitrageurs. This slippage can in principle be eliminated: if an AMM had access to a high-frequency oracle for the CEX price $P_{t}$, the AMM could in principle quote prices arbitrarily close to $P_{t}$, up to the desired asset position $x^{*}\left(P_{t}\right)$.\footnote{A similar design is proposed in \citet{krishnamachari2021dynamic}.} Quoting prices this way would reduce arbitrageurs' profits, allowing the AMM to achieve a payoff arbitrarily close to that of the rebalancing strategy. This design has a number of risks --- it relies heavily on the accuracy of the oracle for $P_{t}$, and leaves open the potential for oracle manipulation --- but in principle an oracle-based AMM could substantially reduce or eliminate $\LVR$.

A related set of AMM designs revolve around selling the right to arbitrage the pool to certain special wallets, and redistributing profits to AMM LPs. This could be done in several ways. Suppose a particular crypto wallet address, which we call the ``authorized participant'' or AP, had the unique right for the first AMM trade in every block. This first execution right effectively conveys the right to arbitrage the pool \citep{mcamm_2022}. Alternatively, the AP could be provided the right to trade with the AMM paying zero fees \citep{pyth_defender_2022}. The AP would then have a large advantage in arbitrage trading, since the AP could profitably trade against arbitrarily small price movements, whereas prices would have to move at least as much as the AMM's percentage trading fees for non-AP wallets to profit from arbitrage trade. The AP would thus be able to capture a large fraction of fees from arbitrage trade. In either of these designs, an AMM protocol aiming to reduce $\LVR$ could thus run an AP wallet itself, doing CEX-DEX arbitrage, and redistributing arbitrage profits to LPs. Alternatively, AMM protocols could simply run periodic auctions --- over longer periods, such as weeks or months --- in which wallets can bid for the right to be authorized participants for some period of time. In principle, potential arbitrageurs should bid the \emph{ex ante} expectation of arbitrage profits, which includes $\LVR$; the protocol can then redistribute these profits to LPs. \citet{adams2024amm} analyzes this design in detail.\footnote{Note that one subtlety about the zero fee design is that, in addition to rebalancing arbitrage profits, APs would be able to capture reversion arbitrage profits; hence, $\LVR$ provides a lower bound for how much revenue wallets could capture from having preferential access to arbitrage the pool.} Both these methods capture either \LVR or its expectation, and redistribute the profits to pool LPs. Another strand of work has argued that \LVR would be reduced or eliminated if orders were sold in \emph{batch auctions}, rather than placed directly in mempools; this design is discussed in \citet{ramseyer2022augmenting} and \citet{canidio2023arbitrageurs}, and related ``request-for-quotes'' mechanisms are now in use in a number of widely used protocols.\footnote{``Batching'' is used in \href{https://swap.cow.fi}{CoWswap}; a number of AMM protocols have also implemented related RFQ-like systems, such as \href{https://blog.uniswap.org/uniswapx-protocol}{Uniswap X}, \href{https://1inch.io/fusion/}{1inch fusion}, and \href{https://developers.paraswap.network/augustusrfq-api-specification}{Paraswap AugustusRFQ}.} We believe that these are interesting directions for future AMM design research.

An earlier line of work seeks to design specific CFMMs with good properties by identifying good bonding functions \citep{port2022mixing,wu2022constant,forgy2021family,krishnamachari2021dynAMMs}. Our work suggests that bonding functions only affect \LVR insofar as they change the marginal liquidity of CFMMs. In certain cases, it may be desirable for marginal liquidity to vary with prices in a systematic way --- this may be the case, for example, for stablecoin swap pairs. In other cases, different bonding functions, or ``universal'' AMM designs such as Uniswap v3, have equivalent infinitesimal \LVR whenever they have equivalent amounts of marginal liquidity for an asset pair.

\section{Conclusion}

\label{sec:conclusion}

Automated market makers are \emph{decentralized} market makers. Market participants can both trade with AMMs, and contribute capital to participate in the liquidity provision activities of the AMM. In the second case, the AMM behaves essentially as an investment, similar to mutual funds or ETFs in traditional financial markets. In this paper, we have constructed a framework for understanding the returns to investments in AMM LP positions. We provide a simple model, in which LP \pnl cleanly decomposes into a ``beta-like'' component reflecting market risk, and an ``alpha-like'' component reflecting the microstructural forces of fee revenue versus adverse selection losses. These components are straightforward to separate empirically: the beta component can be eliminated simply by subtracting the profits of the rebalancing strategy from LP \pnl. Our results have implications for how to measure the returns to LP positions as investments.


{\small\singlespacing
  \bibliographystyle{plainnat}
  \bibliography{references}

\begin{thebibliography}{70}
\providecommand{\natexlab}[1]{#1}
\providecommand{\url}[1]{\texttt{#1}}
\expandafter\ifx\csname urlstyle\endcsname\relax
  \providecommand{\doi}[1]{doi: #1}\else
  \providecommand{\doi}{doi: \begingroup \urlstyle{rm}\Url}\fi

\bibitem[Adams et~al.(2024)Adams, Moallemi, Reynolds, and
  Robinson]{adams2024amm}
Austin Adams, Ciamac Moallemi, Sara Reynolds, and Dan Robinson.
\newblock am-amm: An auction-managed automated market maker.
\newblock \emph{arXiv preprint arXiv:2403.03367}, 2024.

\bibitem[Adams et~al.(2021)Adams, Zinsmeister, Salem, Keefer, and
  Robinson]{adams2021uniswap}
Hayden Adams, Noah Zinsmeister, Moody Salem, River Keefer, and Dan Robinson.
\newblock Uniswap v3 core, 2021.

\bibitem[{Algebra Protocol}(2023)]{algebradexengine}
{Algebra Protocol}.
\newblock {ALGEBRA Ecosystem: Decentralized exchange}, 2023.
\newblock URL
  \url{https://algebra.finance/static/Algerbra%20Tech%20Paper-15411d15f8653a81d5f7f574bfe655ad.pdf}.

\bibitem[Angeris and Chitra(2020)]{angeris2020improved}
Guillermo Angeris and Tarun Chitra.
\newblock Improved price oracles: Constant function market makers.
\newblock In \emph{Proceedings of the 2nd ACM Conference on Advances in
  Financial Technologies}, pages 80--91, 2020.

\bibitem[Angeris et~al.(2019)Angeris, Kao, Chiang, Noyes, and
  Chitra]{angeris2019analysis}
Guillermo Angeris, Hsien-Tang Kao, Rei Chiang, Charlie Noyes, and Tarun Chitra.
\newblock An analysis of uniswap markets.
\newblock \emph{arXiv preprint arXiv:1911.03380}, 2019.

\bibitem[Angeris et~al.(2021{\natexlab{a}})Angeris, Evans, and
  Chitra]{angeris2021replicatingmarketmakers}
Guillermo Angeris, Alex Evans, and Tarun Chitra.
\newblock Replicating market makers.
\newblock \emph{arXiv preprint arXiv:2103.14769}, 2021{\natexlab{a}}.

\bibitem[Angeris et~al.(2021{\natexlab{b}})Angeris, Evans, and
  Chitra]{angeris2021replicatingmonotonicpayoffs}
Guillermo Angeris, Alex Evans, and Tarun Chitra.
\newblock Replicating monotonic payoffs without oracles.
\newblock \emph{arXiv preprint arXiv:2111.13740}, 2021{\natexlab{b}}.

\bibitem[{Aori Protocol}(2023)]{aori}
{Aori Protocol}.
\newblock {Trading} on {Aori}, 2023.
\newblock URL \url{https://www.aori.io/docs/trading}.

\bibitem[Aoyagi(2020)]{aoyagi2020liquidity}
Jun Aoyagi.
\newblock Liquidity provision by automated market makers.
\newblock \emph{SSRN 3674178}, 2020.

\bibitem[Aoyagi and Ito(2021)]{aoyagi2021coexisting}
Jun Aoyagi and Yuki Ito.
\newblock Coexisting exchange platforms: {Limit} order books and automated
  market makers.
\newblock \emph{SSRN 3808755}, 2021.

\bibitem[Aquilina et~al.(2024)Aquilina, Foley, Gambacorta, and
  Krekel]{aquilina2024decentralised}
Matteo Aquilina, Sean Foley, Leonardo Gambacorta, and William Krekel.
\newblock Decentralised dealers? examining liquidity provision in decentralised
  exchanges.
\newblock Technical report, Bank for International Settlements, 2024.

\bibitem[Augustin et~al.(2024)Augustin, Chen-Zhang, and
  Shin]{augustin2024reaching}
Patrick Augustin, Roy Chen-Zhang, and Donghwa Shin.
\newblock Reaching for yield in decentralized financial markets.
\newblock Working paper, 2024.

\bibitem[Barbon and Ranaldo(2021)]{barbon2021quality}
Andrea Barbon and Angelo Ranaldo.
\newblock On the quality of cryptocurrency markets: Centralized versus
  decentralized exchanges.
\newblock \emph{arXiv preprint arXiv:2112.07386}, 2021.

\bibitem[Bertsekas(1971)]{bertsekas1971control}
Dimitri~P Bertsekas.
\newblock \emph{Control of uncertain systems with a set-membership description
  of the uncertainty.}
\newblock PhD thesis, Massachusetts Institute of Technology, 1971.

\bibitem[Black and Scholes(1973)]{black1973pricing}
Fischer Black and Myron Scholes.
\newblock The pricing of options and corporate liabilities.
\newblock \emph{Journal of political economy}, 81\penalty0 (3):\penalty0
  637--654, 1973.

\bibitem[Boueri(2021)]{boueri2021g3m}
Nassib Boueri.
\newblock G3m impermanent loss dynamics.
\newblock \emph{arXiv preprint arXiv:2108.06593}, 2021.

\bibitem[Brolley and Zoican(2023)]{brolley2023demand}
Michael Brolley and Marius Zoican.
\newblock On-demand fast trading on decentralized exchanges.
\newblock \emph{Finance Research Letters}, 51:\penalty0 103350, 2023.

\bibitem[Budish et~al.(2015)Budish, Cramton, and Shim]{budish2015high}
Eric Budish, Peter Cramton, and John Shim.
\newblock The high-frequency trading arms race: Frequent batch auctions as a
  market design response.
\newblock \emph{The Quarterly Journal of Economics}, 130\penalty0 (4):\penalty0
  1547--1621, 2015.

\bibitem[Buterin(2016)]{vbuterin_lets_2016}
Vitalik Buterin.
\newblock Let's run on-chain decentralized exchanges the way we run prediction
  markets, October 2016.
\newblock URL
  \url{www.reddit.com/r/ethereum/comments/55m04x/lets_run_onchain_decentralized_exchanges_the_way/}.

\bibitem[Canidio and Fritsch(2023)]{canidio2023arbitrageurs}
Andrea Canidio and Robin Fritsch.
\newblock Arbitrageurs' profits, lvr, and sandwich attacks: batch trading as an
  amm design response.
\newblock \emph{arXiv preprint arXiv:2307.02074}, 2023.

\bibitem[Capponi and Jia(2021)]{capponi2021adoption}
Agostino Capponi and Ruizhe Jia.
\newblock The adoption of blockchain-based decentralized exchanges.
\newblock \emph{arXiv preprint arXiv:2103.08842}, 2021.

\bibitem[Carr and Lee(2009)]{carr2009volatility}
Peter Carr and Roger Lee.
\newblock Volatility derivatives.
\newblock \emph{Annu. Rev. Financ. Econ.}, 1\penalty0 (1):\penalty0 319--339,
  2009.

\bibitem[Carr and Madan(2001)]{carr2001towards}
Peter Carr and Dilip Madan.
\newblock Towards a theory of volatility trading.
\newblock \emph{Option pricing, interest rates and risk management, handbooks
  in mathematical finance}, 22\penalty0 (7):\penalty0 458--476, 2001.

\bibitem[Carr and Jarrow(1990)]{carr1990stop}
Peter~P Carr and Robert~A Jarrow.
\newblock The stop-loss start-gain paradox and option valuation: A new
  decomposition into intrinsic and time value.
\newblock \emph{The review of financial studies}, 3\penalty0 (3):\penalty0
  469--492, 1990.

\bibitem[Cartea et~al.(2022)Cartea, Drissi, and Monga]{cartea2022decentralised}
{\'A}lvaro Cartea, Fay{\c{c}}al Drissi, and Marcello Monga.
\newblock Decentralised finance and automated market making: Predictable loss
  and optimal liquidity provision.
\newblock \emph{Available at SSRN 4273989}, 2022.

\bibitem[Cartea et~al.(2023)Cartea, Drissi, and Monga]{cartea2023predictable}
{\'A}lvaro Cartea, Fay{\c{c}}al Drissi, and Marcello Monga.
\newblock Predictable losses of liquidity provision in constant function
  markets and concentrated liquidity markets.
\newblock \emph{Applied Mathematical Finance}, 30\penalty0 (2):\penalty0
  69--93, 2023.

\bibitem[{Cata Labs}(2023)]{catalabs2023}
{Cata Labs}.
\newblock Optimising lp performance part 2: Dynamic fees, Nov 2023.
\newblock URL
  \url{https://blog.catalyst.exchange/optimising-lp-performance-part-2-dynamic-fees/}.

\bibitem[Chen and Pennock(2007)]{chenpennock2007}
Y.~Chen and D.M Pennock.
\newblock A utility framework for bounded-loss market makers.
\newblock In \emph{Proceedings of the 23rd Conference on Uncertainty in
  Artificial Intelligence (UAI 2007}, pages 49--56, Vancouver, BC, Canada,
  2007.

\bibitem[Clark(2020)]{clark2020replicating}
Joseph Clark.
\newblock The replicating portfolio of a constant product market.
\newblock \emph{Available at SSRN 3550601}, 2020.

\bibitem[Cox et~al.(1979)Cox, Ross, and Rubinstein]{cox1979option}
John~C Cox, Stephen~A Ross, and Mark Rubinstein.
\newblock Option pricing: A simplified approach.
\newblock \emph{Journal of Financial Economics}, 7\penalty0 (3):\penalty0
  229--263, 1979.

\bibitem[CrocSwap(2022)]{crocswap_usage_2022}
CrocSwap.
\newblock Usage of markout to calculate lp profitability in uniswap v3, 2022.
\newblock URL
  \url{https://crocswap.medium.com/usage-of-markout-to-calculate-lp-profitability-in-uniswap-v3-e32773b1a88e}.

\bibitem[Daian et~al.(2020)Daian, Goldfeder, Kell, Li, Zhao, Bentov,
  Breidenbach, and Juels]{daian2020flash}
Philip Daian, Steven Goldfeder, Tyler Kell, Yunqi Li, Xueyuan Zhao, Iddo
  Bentov, Lorenz Breidenbach, and Ari Juels.
\newblock Flash boys 2.0: Frontrunning in decentralized exchanges, miner
  extractable value, and consensus instability.
\newblock In \emph{2020 IEEE Symposium on Security and Privacy (SP)}, pages
  910--927. IEEE, 2020.

\bibitem[Deng et~al.(2023)Deng, Zong, and Wang]{deng2023static}
Jun Deng, Hua Zong, and Yun Wang.
\newblock Static replication of impermanent loss for concentrated liquidity
  provision in decentralised markets.
\newblock \emph{Operations Research Letters}, 51\penalty0 (3):\penalty0
  206--211, 2023.

\bibitem[Dixit and Pindyck(1994)]{dixit1994investment}
Avinash~K Dixit and Robert~S Pindyck.
\newblock \emph{Investment under uncertainty}.
\newblock Princeton university press, 1994.

\bibitem[Elsts(2023)]{atis2023}
Atis Elsts.
\newblock Conceptualizing {Uniswap} v3 {LP} profit and loss, January 2023.
\newblock URL
  \url{https://atise.medium.com/conceptualizing-uniswap-v3-lp-profit-and-loss-ecbae6e09644}.

\bibitem[Evans(2020)]{evans2020liquidity}
Alex Evans.
\newblock Liquidity provider returns in geometric mean markets.
\newblock \emph{arXiv preprint arXiv:2006.08806}, 2020.

\bibitem[Fang(2024)]{fang2024learning}
Chuck Fang.
\newblock Learning financial innovations: Experience from automated market
  makers.
\newblock \emph{Jacobs Levy Equity Management Center for Quantitative Financial
  Research Paper}, 2024.

\bibitem[{Fenbushi Capital}(2023)]{fenbushi2023}
{Fenbushi Capital}.
\newblock {Ending LP's Losing Game: Exploring the Loss-Versus-Rebalancing (LVR)
  Problem and its Solutions}, Oct 2023.
\newblock URL
  \url{https://mirror.xyz/0xbdA5bCe76bF62d97D9C9dF0425CC10079Df1aD75/bWOCccjVC7eoYKOzgmjXFdhWDc8rrUL6Yei-eugF52s}.

\bibitem[Foley et~al.(2023)Foley, O'Neill, and
  Putni{\c{n}}{\v{s}}]{foley2023better}
Sean Foley, Peter O'Neill, and T{\=a}lis~J Putni{\c{n}}{\v{s}}.
\newblock A better market design? applying ‘automated market makers’ to
  traditional financial markets.
\newblock \emph{Available at SSRN 4459924}, 2023.

\bibitem[Forgy and Lau(2021)]{forgy2021family}
Eric Forgy and Leo Lau.
\newblock A family of multi-asset automated market makers.
\newblock \emph{arXiv preprint arXiv:2111.08115}, 2021.

\bibitem[Fukasawa et~al.(2022)Fukasawa, Maire, and
  Wunsch]{fukasawa2022weighted}
Masaaki Fukasawa, Basile Maire, and Marcus Wunsch.
\newblock Weighted variance swaps hedge against impermanent loss.
\newblock \emph{Available at SSRN 4095029}, 2022.

\bibitem[Glosten and Milgrom(1985)]{glosten1985bid}
Lawrence~R Glosten and Paul~R Milgrom.
\newblock Bid, ask and transaction prices in a specialist market with
  heterogeneously informed traders.
\newblock \emph{Journal of financial economics}, 14\penalty0 (1):\penalty0
  71--100, 1985.

\bibitem[Hasbrouck et~al.(2022)Hasbrouck, Rivera, and Saleh]{hasbrouck2022need}
Joel Hasbrouck, Thomas~J Rivera, and Fahad Saleh.
\newblock The need for fees at a dex: How increases in fees can increase dex
  trading volume.
\newblock \emph{Available at SSRN}, 2022.

\bibitem[Hasbrouck et~al.(2025)Hasbrouck, Rivera, and
  Saleh]{hasbrouck2023economic}
Joel Hasbrouck, Thomas~J Rivera, and Fahad Saleh.
\newblock An economic model of a decentralized exchange with concentrated
  liquidity.
\newblock \emph{Management Science}, 2025.

\bibitem[Heimbach et~al.(2022)Heimbach, Schertenleib, and
  Wattenhofer]{heimbach2022risks}
Lioba Heimbach, Eric Schertenleib, and Roger Wattenhofer.
\newblock Risks and returns of uniswap v3 liquidity providers.
\newblock In \emph{Proceedings of the 4th ACM Conference on Advances in
  Financial Technologies}, pages 89--101, 2022.

\bibitem[Huang and Stoll(1996)]{huang1996dealer}
Roger~D. Huang and Hans~R. Stoll.
\newblock Dealer versus auction markets: A paired comparison of execution costs
  on {NASDAQ} and the {NYSE}.
\newblock \emph{Journal of Financial Economics}, 41\penalty0 (3):\penalty0
  313--357, 1996.

\bibitem[Josojo(2022)]{mcamm_2022}
Josojo.
\newblock {MEV} capturing {AMM} {(McAMM)}, 2022.
\newblock URL \url{https://ethresear.ch/t/mev-capturing-amm-mcamm/13336}.

\bibitem[{Jump Crypto}(2022)]{pyth_defender_2022}
{Jump Crypto}.
\newblock Pyth {Defender} white paper draft.
\newblock Working paper, September 2022.

\bibitem[Klages-Mundt and Schuldenzucker(2022)]{gyroscope2023}
Ariah Klages-Mundt and Steffen Schuldenzucker.
\newblock {Elliptic Concentrated Liquidity Pool (E-CLP): Technical Overview},
  2022.

\bibitem[Krishnamachari et~al.(2021{\natexlab{a}})Krishnamachari, Feng, and
  Grippo]{krishnamachari2021dynAMMs}
Bhaskar Krishnamachari, Qi~Feng, and Eugenio Grippo.
\newblock Dynamic automated market makers for decentralized cryptocurrency
  exchange.
\newblock In \emph{2021 IEEE International Conference on Blockchain and
  Cryptocurrency (ICBC)}, pages 1--2, 2021{\natexlab{a}}.
\newblock \doi{10.1109/ICBC51069.2021.9461100}.

\bibitem[Krishnamachari et~al.(2021{\natexlab{b}})Krishnamachari, Feng, and
  Grippo]{krishnamachari2021dynamic}
Bhaskar Krishnamachari, Qi~Feng, and Eugenio Grippo.
\newblock Dynamic curves for decentralized autonomous cryptocurrency exchanges.
\newblock \emph{arXiv preprint arXiv:2101.02778}, 2021{\natexlab{b}}.

\bibitem[Lambert(2022)]{lambert2022}
Guillaume Lambert.
\newblock Website at \url{https://lambert-guillaume.medium.com/}, 2022.

\bibitem[Lehar and Parlour(2021)]{lehar2021decentralized}
Alfred Lehar and Christine~A Parlour.
\newblock Decentralized exchanges.
\newblock Technical report, Working paper, 2021.

\bibitem[Lehar et~al.(2024)Lehar, Parlour, and Zoican]{lehar2022liquidity}
Alfred Lehar, Christine~A Parlour, and Marius Zoican.
\newblock Liquidity fragmentation on decentralized exchanges.
\newblock \emph{Available at SSRN 4267429}, 2024.

\bibitem[Liao and Robinson(2022)]{liao2022dominance}
Gordon Liao and Dan Robinson.
\newblock The dominance of uniswap v3 liquidity.
\newblock \url{https://blog.uniswap.org/uniswap-v3-dominance}, 2022.
\newblock Accessed: 2024-05-01.

\bibitem[Lintner(1965)]{Lintner1965}
John Lintner.
\newblock The valuation of risk assets and the selection of risky investments
  in stock portfolios and capital budgets.
\newblock \emph{The Review of Economics and Statistics}, 47\penalty0
  (1):\penalty0 13--37, February 1965.
\newblock \doi{10.2307/1924119}.

\bibitem[Lu and K\"oppelmann(2017)]{lu_building_2017}
Alan Lu and Martin K\"oppelmann.
\newblock Building a {Decentralized} {Exchange} in {Ethereum}, March 2017.
\newblock URL
  \url{https://blog.gnosis.pm/building-a-decentralized-exchange-in-ethereum-eea4e7452d6e}.

\bibitem[Milionis et~al.(2023{\natexlab{a}})Milionis, Moallemi, and
  Roughgarden]{2023-02-lvr-fee-model}
Jason Milionis, Ciamac~C Moallemi, and Tim Roughgarden.
\newblock Automated market making and arbitrage profits in the presence of
  fees.
\newblock \emph{arXiv preprint arXiv:2305.14604}, 2023{\natexlab{a}}.

\bibitem[Milionis et~al.(2023{\natexlab{b}})Milionis, Moallemi, and
  Roughgarden]{jason_exchange_complexity}
Jason Milionis, Ciamac~C. Moallemi, and Tim Roughgarden.
\newblock {Complexity-Approximation Trade-offs in Exchange Mechanisms: AMMs vs.
  LOBs}.
\newblock In \emph{Financial Cryptography and Data Security}. Springer
  International Publishing, 2023{\natexlab{b}}.
\newblock ISBN 978-3-031-47753-9.

\bibitem[Milionis et~al.(2024)Milionis, Moallemi, and
  Roughgarden]{milionis2023myersonian}
Jason Milionis, Ciamac~C. Moallemi, and Tim Roughgarden.
\newblock {A Myersonian Framework for Optimal Liquidity Provision in Automated
  Market Makers}.
\newblock In \emph{15th Innovations in Theoretical Computer Science Conference
  (ITCS 2024)}, Leibniz International Proceedings in Informatics (LIPIcs),
  Dagstuhl, Germany, 2024. Schloss Dagstuhl -- Leibniz-Zentrum f{\"u}r
  Informatik.

\bibitem[O'Neill(2022)]{oneill2022}
Peter O'Neill.
\newblock {Can Markets be Fully Automated? Evidence From an ``Automated Market
  Maker''}, January 2022.
\newblock URL
  \url{https://raw.githubusercontent.com/petero1111/website/gh-pages/ONeill_JMP_2022.pdf}.

\bibitem[Park(2021)]{park2021conceptual}
Andreas Park.
\newblock The conceptual flaws of constant product automated market making.
\newblock \emph{Available at SSRN 3805750}, 2021.

\bibitem[Pennock and Sami(2007)]{pennock2007computational}
David~M Pennock and Rahul Sami.
\newblock Computational aspects of prediction markets.
\newblock \emph{Algorithmic game theory}, pages 651--674, 2007.

\bibitem[Port and Tiruviluamala(2022)]{port2022mixing}
Alexander Port and Neelesh Tiruviluamala.
\newblock Mixing constant sum and constant product market makers.
\newblock \emph{arXiv preprint arXiv:2203.12123}, 2022.

\bibitem[Ramseyer et~al.(2022)Ramseyer, Goyal, Goel, and
  Mazi{\`e}res]{ramseyer2022augmenting}
Geoffrey Ramseyer, Mohak Goyal, Ashish Goel, and David Mazi{\`e}res.
\newblock Augmenting batch exchanges with constant function market makers.
\newblock \emph{arXiv preprint arXiv:2210.04929}, 2022.

\bibitem[Ross(1976)]{ross1976arbitrage}
Stephen~A Ross.
\newblock The arbitrage theory of capital asset pricing.
\newblock \emph{Journal of Economic Theory}, 13\penalty0 (3):\penalty0
  341--360, 1976.

\bibitem[Sharpe(1964)]{sharpe1964capital}
William~F Sharpe.
\newblock Capital asset prices: A theory of market equilibrium under conditions
  of risk.
\newblock \emph{The journal of finance}, 19\penalty0 (3):\penalty0 425--442,
  1964.

\bibitem[Simon et~al.(1994)Simon, Blume, et~al.]{simon1994mathematics}
Carl~P Simon, Lawrence Blume, et~al.
\newblock \emph{Mathematics for economists}, volume~7.
\newblock Norton New York, 1994.

\bibitem[Tassy and White(2020)]{tassy2020growth}
Martin Tassy and David White.
\newblock Growth rate of a liquidity provider’s wealth in $xy= c$ automated
  market makers, 2020.

\bibitem[Wu and McTighe(2022)]{wu2022constant}
Mike Wu and Will McTighe.
\newblock Constant power root market makers.
\newblock \emph{arXiv preprint arXiv:2205.07452}, 2022.

\end{thebibliography}
}

\clearpage

\appendix
\crefalias{section}{appendix}

\begin{center}
{\LARGE{}Appendix}
\par\end{center}

\section{Proofs}

\label{sec:proofs}

\subsection{Proof of \Cref{le:envelope}}

The first part follows from the fact that $\Cscr\subset\R_{+}^{2}$ and $P\geq0$. The second part is the envelope theorem or Danskin's theorem \citep{bertsekas1971control}. The third part follows from the concavity of $V(\cdot)$, as a pointwise minimum of a collection of affine functions. 

\subsection{Proof of \Cref{lem:lvr}}

First, we show that $\LVR_{t}$ is equal to expression \eqref{eq:lvr}. The smoothness condition of \Cref{as:smooth-v} Part~\ref{pt:smooth-v} allows us to apply It\^o's lemma 
 to $V(\cdot)$ to obtain 
\begin{equation}
\begin{split}dV_{t} & =V'(P_{t})\,dP_{t}+\tfrac{1}{2}V''(P_{t})\left(dP_{t}\right)^{2}\\
 & =V'(P_{t})\,dP_{t}+\tfrac{1}{2}V''(P_{t})\,\sigma^{2}P_{t}^{2}\,dt\\
 & =x^{*}(P_{t})\,dP_{t}+\tfrac{1}{2}V''(P_{t})\,\sigma^{2}P_{t}^{2}\,dt,
\end{split}
\label{eq:ito}
\end{equation}
where the last step follows from \Cref{le:envelope} Part~\ref{pt:Vp}. Comparing with \eqref{eq:rebalance}, we obtain \eqref{eq:lvr}. Finally, the fact that $\ell(\sigma,P)\geq0$ follows from \Cref{le:envelope} Part~\ref{pt:Vpp}.

Next, we show that the cumulative profits of rebalancing arbitrageurs over the time interval $[0,T]$ is equal to $\LVR_{T}$. We start with a discrete approximation to the arbitrage profit, indexed by $N\geq1$. Suppose arbitrageurs arrive sequentially, so that the $i$th arbitrageur arrives at time $\tau_{i}$, for $1\leq i\leq N$. For convenience, set $\tau_{0}\defeq0$ and $\tau_{N+1}\defeq T$. For each $1\leq i\leq N$, at time $\tau_{i}$, the $i$th arbitrageur observes the price $P_{\tau_{i}}$, rebalances the pool from $\big(x^{*}(P_{\tau_{i-1}}),y^{*}(P_{\tau_{i-1}})\big)$ to $\big(x^{*}(P_{\tau_{i}}),y^{*}(P_{\tau_{i}})\big)$. In other words, the arbitrageur purchases $x^{*}(P_{\tau_{i-1}})-x^{*}(P_{\tau_{i}})$ units of the risky asset from the CFMM at average price : 
\[
P_{i}^{\CFMM}\defeq-\frac{y^{*}(P_{\tau_{i}})-y^{*}(P_{\tau_{i-1}})}{x^{*}(P_{\tau_{i}})-x^{*}(P_{\tau_{i-1}})}.
\]
The arbitrageur can then sell these units on the external market at price $P_{\tau_{i}}$ and earn profits (in the num\'eraire) from the difference in price according to 
\[
\delimitershortfall=-1pt\left(P_{\tau_{i}}-P_{i}^{\CFMM}\right)\left[x^{*}(P_{\tau_{i-1}})-x^{*}(P_{\tau_{i}})\right]=P_{\tau_{i}}\left[x^{*}(P_{\tau_{i-1}})-x^{*}(P_{\tau_{i}})\right]+\left[y^{*}(P_{\tau_{i-1}})-y^{*}(P_{\tau_{i}})\right].
\]
Denote by $\ARB_{T}^{(N)}$ the aggregate arbitrage profits. Summing over $1\leq i\leq N$, telescoping the sum, and applying summation-by-parts yields 
\[
\delimitershortfall=-1pt\begin{split}\ARB_{T}^{(N)} & \defeq\sum_{i=1}^{N}\left\{ P_{\tau_{i}}\left[x^{*}(P_{\tau_{i-1}})-x^{*}(P_{\tau_{i}})\right]+\left[y^{*}(P_{\tau_{i-1}})-y^{*}(P_{\tau_{i}})\right]\right\} \\
 & =\sum_{i=1}^{N}P_{\tau_{i}}\left[x^{*}(P_{\tau_{i-1}})-x^{*}(P_{\tau_{i}})\right]+y^{*}(P_{0})-y^{*}(P_{\tau_{N}})\\
 & =P_{0}x^{*}(P_{0})+y^{*}(P_{0})+\sum_{i=0}^{N}x^{*}(P_{\tau_{i}})\left[P_{\tau_{i+1}}-P_{\tau_{i}}\right]-P_{T}x^{*}(P_{\tau_{N}})-y^{*}(P_{\tau_{N}}).
\end{split}
\]
Observe that the sum in the final expression is the discrete approximation of an It\^o integral. Assume that the time partition mesh over $[0,T]$ shrinks to zero as $N\tends\infty$. Taking the limit as $N\tends\infty$ and passing to continuous time, the sum converges to an It\^o integral, which is well-defined under \Cref{as:smooth-v} Part~\ref{pt:bounded}. Further, $\tau_{N}\tends T$, so that $P_{\tau_{N}}\tends P_{T}$, and $x^{*}(P_{\tau_{N}})\tends x^{*}(P_{T})$, $y^{*}(P_{\tau_{N}})\tends y^{*}(P_{T})$. Thus, it holds that 
\[
\ARB_{T}\defeq\lim_{N\tends\infty}\ARB_{T}^{(N)}=V(P_{0})+\int_{0}^{T}x^{*}(P_{t})\,dP_{t}-V(P_{T}).\qedhere
\]
Hence, the cumulative profits of rebalancing arbitrageurs from time $0$ to time $T$ are equal to \LVR, defined in \eqref{eq:lvrdef}.

\subsection{LVR, Marginal Liquidity, and Bonding Function Curvature}

\label{subsec:dxdpformula}

For sufficiently smooth CFMM bonding functions, the marginal liquidity can be expressed in terms of derivatives of the CFMM bonding function:

\begin{equation}
\frac{dx}{dP}=\frac{\frac{\partial f}{\partial y}}{\left(\frac{\partial^{2}f}{\partial x^{2}}+P^{2}\frac{\partial^{2}f}{\partial y^{2}}-2P\frac{\partial^{2}f}{\partial x\partial y}\right)}\label{eq:dxdPbigexp}
\end{equation}

Qualitatively, \eqref{eq:dxdPbigexp} implies that marginal liquidity, $x^{\prime}\left(P\right)$, is related to the curvature of the CFMM invariant curves. The denominator of \eqref{eq:dxdPbigexp} is equal to $P^{2}$ times the negative of the determinant of the bordered Hessian of $f$. $f$ is strictly quasiconcave --- that is, the upper level sets of $f$ are convex --- if and only if this determinant is positive; moreover, the magnitude of the determinant is related to the curvature of the level curves of $f$ \citep[p. 542][]{simon1994mathematics}. Thus, CFMM invariants with ``flatter'', more linear level curves will have greater marginal liquidity $\frac{dx}{dP}$, and also greater $\LVR$.

\eqref{eq:dxdPbigexp} is also useful because it can be used to calculate $\frac{dx}{dP}$ using analytic expressions for bonding functions, which may be useful for computing $\LVR$ in practice.

\subsubsection{Derivation of \eqref{eq:dxdPbigexp}}


The Lagrangian of the pool expenditure minimization problem, \eqref{eq:pool-min}, is: 
\[
\Lambda=Px+y+\lambda\left[f\left(x,y\right)-L\right]
\]
The optimal solution is characterized by the FOCs: 
\begin{equation}
\frac{\partial\Lambda}{\partial x}:\ P+\lambda\frac{\partial f}{\partial x}=0\label{eq:LambdaXfoc}
\end{equation}
\begin{equation}
\frac{\partial\Lambda}{\partial y}:\ 1+\lambda\frac{\partial f}{\partial y}=0\label{eq:LambdaYfoc}
\end{equation}
\begin{equation}
\frac{\partial\Lambda}{\partial\lambda}:\ f\left(x,y\right)-L=0\label{eq:Lambdalamfoc}
\end{equation}
Now, we will take $\frac{dx}{dP}$ by applying the implicit function theorem to this system of first-order conditions. The derivatives of the FOCs are:

\[
\frac{\partial}{\partial P}\frac{\partial\Lambda}{\partial x}:\ 1
\]
\[
\frac{\partial}{\partial P}\frac{\partial\Lambda}{\partial y}:\ 0
\]
\[
\frac{\partial}{\partial P}\frac{\partial\Lambda}{\partial\lambda}:\ 0
\]

\[
\frac{\partial}{\partial x}\frac{\partial\Lambda}{\partial x}:\ \lambda\frac{\partial^{2}f}{\partial x^{2}}
\]
\[
\frac{\partial}{\partial x}\frac{\partial\Lambda}{\partial y}:\ \lambda\frac{\partial^{2}f}{\partial x\partial y}
\]
\[
\frac{\partial}{\partial x}\frac{\partial\Lambda}{\partial\lambda}:\ \frac{\partial f}{\partial x}
\]

\[
\frac{\partial}{\partial y}\frac{\partial\Lambda}{\partial x}:\ \lambda\frac{\partial^{2}f}{\partial x\partial y}
\]
\[
\frac{\partial}{\partial y}\frac{\partial\Lambda}{\partial y}:\ \lambda\frac{\partial^{2}f}{\partial y^{2}}
\]
\[
\frac{\partial}{\partial y}\frac{\partial\Lambda}{\partial\lambda}:\ \frac{\partial f}{\partial y}
\]

\[
\frac{\partial}{\partial\lambda}\frac{\partial\Lambda}{\partial x}:\ -\frac{\partial f}{\partial x}
\]
\[
\frac{\partial}{\partial\lambda}\frac{\partial\Lambda}{\partial y}:\ -\frac{\partial f}{\partial y}
\]
\[
\frac{\partial}{\partial\lambda}\frac{\partial\Lambda}{\partial\lambda}:\ 0
\]
Hence, we wish to solve the following system of equations for $\frac{dx}{dP}$: 
\[
\frac{\partial}{\partial P}\frac{\partial\Lambda}{\partial x}+\frac{\partial}{\partial x}\frac{\partial\Lambda}{\partial x}\frac{dx}{dP}+\frac{\partial}{\partial y}\frac{\partial\Lambda}{\partial x}\frac{dy}{dP}+\frac{\partial}{\partial\lambda}\frac{\partial\Lambda}{\partial x}\frac{d\lambda}{dP}=0
\]
\[
\frac{\partial}{\partial P}\frac{\partial\Lambda}{\partial y}+\frac{\partial}{\partial x}\frac{\partial\Lambda}{\partial y}\frac{dx}{dP}+\frac{\partial}{\partial y}\frac{\partial\Lambda}{\partial y}\frac{dy}{dP}+\frac{\partial}{\partial\lambda}\frac{\partial\Lambda}{\partial y}\frac{d\lambda}{dP}=0
\]
\[
\frac{\partial}{\partial P}\frac{\partial\Lambda}{\partial\lambda}+\frac{\partial}{\partial x}\frac{\partial\Lambda}{\partial\lambda}\frac{dx}{dP}+\frac{\partial}{\partial y}\frac{\partial\Lambda}{\partial\lambda}\frac{dy}{dP}+\frac{\partial}{\partial\lambda}\frac{\partial\Lambda}{\partial\lambda}\frac{d\lambda}{dP}=0
\]
Substituting, we have:

\begin{equation}
1+\lambda\frac{\partial^{2}f}{\partial x^{2}}\frac{dx}{dP}+\lambda\frac{\partial^{2}f}{\partial x\partial y}\frac{dy}{dP}-\frac{\partial f}{\partial x}\frac{d\lambda}{dP}=0\label{eq:csFOCtemp1}
\end{equation}
\begin{equation}
0+\lambda\frac{\partial^{2}f}{\partial x\partial y}\frac{dx}{dP}+\lambda\frac{\partial^{2}f}{\partial y^{2}}\frac{dy}{dP}-\frac{\partial f}{\partial y}\frac{d\lambda}{dP}=0\label{eq:csFOCtemp2}
\end{equation}
\begin{equation}
\frac{\partial f}{\partial x}\frac{dx}{dP}+\frac{\partial f}{\partial y}\frac{dy}{dP}=0\label{eq:csFOCtemp3}
\end{equation}
Now, applying \eqref{eq:LambdaXfoc} and \eqref{eq:LambdaYfoc}, we have: 
\begin{equation}
\frac{\partial f}{\partial x}=P\frac{\partial f}{\partial y}\label{eq:csFOCtemp4}
\end{equation}
Thus, we can simplify \eqref{eq:csFOCtemp3} to: 
\[
\frac{dy}{dP}=-P\frac{dx}{dP}
\]
Substituting into \eqref{eq:csFOCtemp1} and \eqref{eq:csFOCtemp2}, we have: 
\[
1+\lambda\frac{\partial^{2}f}{\partial x^{2}}\frac{dx}{dP}+\lambda\frac{\partial^{2}f}{\partial x\partial y}\left(-P\frac{dx}{dP}\right)-\frac{\partial f}{\partial x}\frac{d\lambda}{dP}=0
\]
\[
0+\lambda\frac{\partial^{2}f}{\partial x\partial y}\frac{dx}{dP}+\lambda\frac{\partial^{2}f}{\partial y^{2}}\left(-P\frac{dx}{dP}\right)-\frac{\partial f}{\partial y}\frac{d\lambda}{dP}=0
\]
Rearranging, 
\begin{equation}
\left(\frac{\partial^{2}f}{\partial x^{2}}-P\frac{\partial^{2}f}{\partial x\partial y}\right)\lambda\frac{dx}{dP}=\frac{\partial f}{\partial x}\frac{d\lambda}{dP}-1\label{eq:csFOCtemp7}
\end{equation}
\begin{equation}
\left(\frac{\partial^{2}f}{\partial x\partial y}-P\frac{\partial^{2}f}{\partial y^{2}}\right)\lambda\frac{dx}{dP}=\frac{\partial f}{\partial y}\frac{d\lambda}{dP}\label{eq:csFOCtemp6}
\end{equation}
Now, multiply \eqref{eq:csFOCtemp6} by $P$, applying \eqref{eq:csFOCtemp4}, and subtract \eqref{eq:csFOCtemp7}, to get: 
\[
\left(\frac{\partial^{2}f}{\partial x^{2}}+P^{2}\frac{\partial^{2}f}{\partial y^{2}}-2P\frac{\partial^{2}f}{\partial x\partial y}\right)\lambda\frac{dx}{dP}=-1
\]
Now, 
\[
\frac{dx}{dP}=\frac{-1}{\lambda\left(\frac{\partial^{2}f}{\partial x^{2}}+P^{2}\frac{\partial^{2}f}{\partial y^{2}}-2P\frac{\partial^{2}f}{\partial x\partial y}\right)}
\]
Now, to solve for $\lambda$, we simply use \eqref{eq:LambdaYfoc}. Hence, we have: 
\begin{equation}
\frac{dx}{dP}=\frac{\frac{\partial f}{\partial y}}{\left(\frac{\partial^{2}f}{\partial x^{2}}+P^{2}\frac{\partial^{2}f}{\partial y^{2}}-2P\frac{\partial^{2}f}{\partial x\partial y}\right)}\label{eq:dxdpbigexpapp}
\end{equation}
This is \eqref{eq:dxdPbigexp}.


\section{Other Results}

\label{sec:otherresults}

\subsection{LVR and Impermanent Loss in a Two-Step Binomial Tree}

\label{subsec:binomtree}

\paraheader{Relationships to option strategies.}
The binomial tree example also helps illustrate the analogy between CFMM LP payoffs and European options. The time-2 difference between the payoffs of the rebalancing strategy and the buy-and-hold strategy is positive when prices mean-revert, and negative when prices diverge. Hence, payoffs are similar to those of a short European straddle or strangle position, which involves selling calls and puts which expire after two periods. The positive payoff when prices revert can be thought of as the option premia collected from selling the straddle, and the negative payoffs when prices diverge can be thought of as the payouts to the option buyer, which are made if either the call or the put sold expire in-the-money. The CFMM LP position has a similar pattern of payoffs, but makes 0 profits if prices end where they started. An CFMM LP position, ignoring fees, can thus be thought of like giving away a straddle position, without collecting any upfront option premia. Viewed this way, the equivalence between the rebalancing strategy and the static European strangle on the binomial tree reflects the classic idea that static option positions can be replicated by dynamically trading the underlying asset; in this case, European straddles and strangles are replicated by a strategy which sells the risky asset when prices increase and buys when prices decrease.



\subsection{Weighted Geometric Mean Market Makers}

\label{sec:wgmm}

Weighted geometric mean market makers have the special property that the instantaneous \LVR per dollar of pool value, i.e., $\ell(\sigma,P)/V(P)$, is a constant. The following theorem establishes that these are essentially the only CFMMs for which this is true: \begin{theorem}\label{th:gmm} Suppose a CFMM satisfies
\begin{equation}
\frac{\ell(\sigma,P)}{V(P)}=c(\sigma),\quad\forall\ P\geq0.\label{eq:ell-v}
\end{equation}
Then, we have
\begin{equation}
V(P)=L_{1}P^{\theta(\sigma)}+L_{2}P^{1-\theta(\sigma)},\label{eq:V-gmm-sym}
\end{equation}
for free constants $L_{1},L_{2}\geq0$, where
\[
\theta(\sigma)\defeq\frac{1-\sqrt{1-8c(\sigma)/\sigma^{2}}}{2}\leq\frac{1}{2}.
\]
\end{theorem}

Comparing with \Cref{ex:gmm}, observe that \eqref{eq:V-gmm-sym} states that the pool can only be the ``sum'' of $\theta$ and $1-\theta$ weighted geometric mean market makers. The two degrees of freedom are intuitive, since $\theta$ and $1-\theta$ are exchangeable in \eqref{eq:const}.

\begin{proof}[\proofnamest{Proof of \Cref{th:gmm}}] We construct the following ODE from the \eqref{eq:ell-v} along with \eqref{eq:ilvr},
\[
P^{2}V''(P)+\bar{c}V(P)=0,
\]
with constant $\bar{c}\defeq2c/\sigma^{2}$. Make the substitution $P=e^{z}$, to arrive at the equivalent ODE,
\[
V''(z)-V'(z)+\bar{c}V(z)=0,
\]
which when solved, along with the known limit condition from \eqref{eq:pool-min} that $V(z)\tends0$ as $z\to-\infty$, by the usual method of linear ODEs results in the generic solution,
\[
V(P)=L_{1}P^{\frac{1-\sqrt{1-4\bar{c}}}{2}}+L_{2}P^{\frac{1+\sqrt{1-4\bar{c}}}{2}}=L_{1}P^{\theta}+L_{2}P^{1-\theta}.
\]
Note that the above calculation is allowed because the quantity under the root is necessarily non-negative, as if it were not, then $V(P)$ would not be everywhere concave, which must be the case by \Cref{le:envelope}. \end{proof}

\subsection{Multi-Dimensional Generalization}

\label{subsec:multidim}

In this section, we describe the multi-dimensional generalization of our results. Specifically, denote by vectors $x\in\R_{+}^{n}$ the reserves in $n\geq2$ assets (none of which need be the num\'eraire), and $P_{t}\in\R_{+}^{n}$ a vector of prices (in terms of the num\'eraire). We assume that the price vector evolves according to geometric Brownian motion, i.e.,
\[
dP_{t}=\diag(P_{t})\Sigma_{t}^{1/2}\,dB_{t}^{\Q},\quad\forall\ t\geq0,
\]
with covariance matrix of returns $\Sigma_{t}\in\R^{n\times n}$, $\Sigma_{t}\succeq0$, and where $B_{t}^{\Q}$ is a standard $\Q$-Brownian motion on $\R^{n}$.

Given a bonding function $f\colon\R_{+}^{n}\rightarrow\R$, define the pool value function $V\colon\R_{+}^{n}\rightarrow\R_{+}$ according to
\[
\begin{array}{lll}
V(P)\defeq & \minimize_{x\in\R_{+}^{n}} & P^{\top}x\\
 & \subjectto & f(x)=L.
\end{array}
\]
Analogous to \Cref{as:smooth-v}, we will assume that an optimal solution $x^{*}(P)$ exists for all $P\in\R_{+}^{n}$, that $V(\cdot)$ is twice continuously differentiable, and a suitable square-integrability condition on $x^{*}(\cdot)$.

Analogous to \Cref{le:envelope}, we have \begin{lemma}\label{le:envelope-md} For all prices $P\in\R_{+}^{n}$, the pool value function satisfies:

\begin{enumerate}
\item \label{pt:V-md} $V(P)\geq0$.
\item \label{pt:Vp-md} $\nabla V(P)=x^{*}(P)\geq0$.
\item \label{pt:Vpp-md} $\nabla^{2}V(P)=\nabla x^{*}(P)\preceq0$.
\end{enumerate}
\end{lemma}

Define the rebalancing strategy by $x_{t}=x^{*}(P_{t})$, with value
\[
R_{t}=V_{0}+\int_{0}^{t}x^{*}(P_{s})^{\top}dP_{s},\quad\forall\ t\geq0.
\]
Then, we have the following multi-dimensional analog of \Cref{lem:lvr}: \begin{theorem}\label{th:lvr-md} Loss-versus-rebalancing takes the form
\[
\LVR_{t}=\int_{0}^{t}\ell(\Sigma_{s},P_{s})\,ds,\quad\forall\ t\geq0,
\]
where we define, for $P\geq0$, the instantaneous \LVR
\[
\ell(\Sigma,P)\defeq-\tfrac{1}{2}\tr\left[\diag(P)\Sigma\diag(P)\,\nabla x^{*}(P)\right]\geq0,
\]
where we have applied \Cref{le:envelope-md}. In the case where $\Sigma=\sigma^{2}I$, i.e., i.i.d.\ assets, we have that
\[
\ell(\Sigma,P)=-\frac{\sigma^{2}}{2}\tr\left[\diag(P)^{2}\,\nabla x^{*}(P)\right]=-\frac{\sigma^{2}}{2}\ \sum_{i=1}^{n}\ P_{i}^{2}\frac{\partial}{\partial P_{i}}x^{*}(P)\geq0.
\]
In particular, \LVR is a non-negative, non-decreasing, and predictable process. \end{theorem} \begin{proof} Applying It\^o's lemma to $V_{t}=V(P)$,
\[
\begin{split}dV_{t} & =\nabla V(P_{t})^{\top}\,dP_{t}+\tfrac{1}{2}(dP_{t})^{\top}\nabla^{2}V(P_{t})\,dP_{t}\\
 & =x^{*}(P_{t})^{\top}\,dP_{t}+\tfrac{1}{2}\tr\left[\Sigma_{t}^{1/2}\diag(P)\nabla^{2}V(P_{t})\diag(P)\Sigma_{t}^{1/2}\right]\,dt\\
 & =dR_{t}-\ell(\Sigma_{t},P_{t})\,dt.
\end{split}
\]
The rest of the result follows as in the proof of \Cref{lem:lvr}. \end{proof}


\section{Data and Measurement}

\label{sec:measurement}

\subsection{Data}
\label{subsec:data}

\paraheader{Prices.} We download minute-level USDC-ETH prices from the
Binance API. We use close prices at the end of each minute for $P_{t}$.

\medskip%
\paraheader{Uniswap.} We download data on the Uniswap~v2 WETH-USDC pool
from Dune Analytics, a data provider which aggregates data from the
Ethereum blockchain into SQL databases. The queries we use to extract
this data are included in \Cref{subsec:sqlqueries}.

\medskip%
\paraheader{Mints and burns.} In each minute, we observe the gross amounts
of each asset in which are withdrawn through ``burns'', and deposited
through ``mints''. Let $(x_{t}^{\text{mint}},y_{t}^{\text{mint}})$ and $(x_{t}^{\text{burn}},y_{t}^{\text{burn}})$
be the total amounts of each asset $x$ and $y$ which are minted
and burned respectively, between time $t-1$ and time $t$. We will
value minted and burned assets at the time-$t$ closing price $P_{t}$.
Thus, the monetary value of mints and burns respectively are:
\[
  \Pi_{t}^{\text{mint}} \defeq
  y_{t}^{\text{mint}} + P_{t}x_{t}^{\text{mint}},
  \quad
  \Pi_{t}^{\text{burn}}\defeq y_{t}^{\text{burn}}+P_{t}x_{t}^{\text{burn}}.
\]
Let $(x_{t},y_{t})$ be the total asset holdings of the pool at time
$t$. As in the model, define the pool value at time $t$ by
\[
V_{t}\defeq y_{t}+P_{t}x_{t}.
\]
We can calculate $\Delta\text{LP\ \ensuremath{\pnl_{t}}}$, the change
in $\pnl$ of the pool, from period $t-1$ to $t$, as
\begin{equation}
\Delta\text{LP\ \ensuremath{\pnl_{t}}}\defeq
V_{t}+\Pi_{t}^{\text{burn}}-\Pi_{t}^{\text{mint}}-V_{t-1}.
\label{eq:empiricsDeltaPnL}
\end{equation}
In words, this is the value of pool reserves at time $t$ valued at
price $P_{t}$, plus burned assets and minus minted assets valued
at $P_{t}$, minus the value of pool reserves at time $t-1$ valued
at $P_{t-1}$. Note that, in contrast to our simplifying assumption
in the model that fees are paid in the num\'eraire, in practice in Uniswap
v2 fees are paid directly into the pool reserves; hence, the pool
$\pnl$ includes transaction fees paid into the pool.

\medskip%
\paraheader{Rebalancing strategy}. We rebalance the pool at different
time frequencies. 
For each rebalancing frequency, we compute
the returns of a strategy which at any point in time holds as much
ETH as the pool holds at the start of the period. 
For example, if the rebalancing frequency is daily, we set $x_{t}^{RB}$
at any minute $t$ equal to the LP pool reserves at the start of the
day containing the minute $t$.
We then calculate the returns on the rebalancing strategy using expression \eqref{eq:empiricsDeltaRBPnL}, that is:
\begin{equation}
\Delta\text{RB\ \ensuremath{\pnl_{t}}}=x_{t}^{RB}\left(P_{t+1}-P_{t}\right).
\label{eq:empiricsDeltaRBPnLapp}
\end{equation}

\medskip%
\paraheader{Fees.} In each minute, we compute the gross amount of each
asset in the pair bought and sold. The Uniswap~v2 pool has a fixed
fee rate of 30bps on the contributed asset; we thus calculate fees in each asset by multiplying
the gross amount contributed of each asset by 0.003. Call these fees
$x_{t}^{fee}$ and $y_{t}^{fee}$ in period $t$. We value fees at the period $t$
price; thus, the monetary value of fees in period $t$, which we will
call $\Delta\FEE_{t}$, is:
\begin{equation}
\Delta\FEE_{t}\defeq y_{t}^{fee}+P_{t}x_{t}^{fee}.
\label{eq:empiricsDeltaFEEapp}
\end{equation}
Note also that Uniswap v2 allows for ``flash loans'', in which assets are withdrawn and returned within a single transaction. Flash loans which are accounted for simply as swaps in Uniswap v2; thus, the calculation in \eqref{eq:empiricsDeltaFEEapp} also correctly accounts for flash loans revenues.\footnote{See the \href{https://docs.uniswap.org/contracts/v2/guides/smart-contract-integration/using-flash-swaps}{Uniswap v2 documentation} of flash swaps.}

\medskip%
\paraheader{LVR.}  We compute a realized daily volatility using USDC-ETH prices from the Binance
API sampled at 60 minute intervals. Let $\Delta \LVR_{t}$ be the increment of \LVR in period
$t$. As in \Cref{ex:cpm}, we then calculate
$\Delta\LVR_{t}$ simply as
\begin{equation}
\Delta\LVR_{t}\defeq\frac{\hat\sigma^2_{t}}{8} \times V_t \times \Delta t,
\label{eq:empiricsDeltaLVRapp}
\end{equation}
where $\hat \sigma_t$ denotes the realized daily volatilty estimate for the day containing period
$t$, and $\Delta t = 1/(24\times 60)$ corresponds to a one minute period. This
is a discrete approximation of \eqref{eq:lvr-cpm}.

\medskip%
\paraheader{Adding everything up.} We then calculate the cumulative returns,
the left side of \eqref{eq:empiricaldecomp}, as:
\[
\text{LP \pnl}_{t}-\int_{0}^{t}x^{*}(P_{s})\,dP_{s}\defeq\sum_{t=1}^{T}\left(\Delta\pnl_{t}-\Delta\text{RB\ \ensuremath{\pnl_{t}}}\right),
\]
using the definitions of $\Delta\text{LP\ \ensuremath{\pnl_{t}}}$ and $\Delta\text{RB\ \ensuremath{\pnl_{t}}}$ in \eqref{eq:empiricsDeltaPnL} and \eqref{eq:empiricsDeltaRBPnLapp} respectively.
Note that different rebalancing frequencies give slightly different values of
for $\Delta\text{RB\ \ensuremath{\pnl_{t}}}$.
We calculate the right side of \eqref{eq:empiricaldecomp} as:
\[
\FEE_{t}-\LVR_{t}\defeq\sum_{t=1}^{T}\left(\Delta\FEE_{t}-\Delta\LVR_{t}\right),
\]
using the definitions of $\Delta\FEE_{t}$ and $\Delta\LVR_{t}$ in \eqref{eq:empiricsDeltaFEEapp} and \eqref{eq:empiricsDeltaLVRapp} respectively.

\subsection{Dune SQL Queries}

\label{subsec:sqlqueries}

This appendix contains the SQL queries we use on Dune to extract Uniswap~v2 WETH-USDC data.

\bigskip

\medskip%
\paraheader{Mints.}
{\singlespacing\small
\begin{lstlisting}[
           language=SQL,
           showspaces=false,
           basicstyle=\ttfamily,
           numbers=left,
           numberstyle=\tiny,
           commentstyle=\color{gray}
        ]
SELECT to_char(evt_block_time, 'YYYY-MM-DD"T"HH24:MI:SSOF') AS ts, *
FROM uniswap_v2."Pair_evt_Mint"
WHERE contract_address = '\xb4e16d0168e52d35cacd2c6185b44281ec28c9dc'
ORDER BY evt_block_number, evt_index ASC
\end{lstlisting}
}

\medskip%
\paraheader{Burns.}
{\singlespacing\small
\begin{lstlisting}[
           language=SQL,
           showspaces=false,
           basicstyle=\ttfamily,
           numbers=left,
           numberstyle=\tiny,
           commentstyle=\color{gray}
        ]
SELECT to_char(evt_block_time, 'YYYY-MM-DD"T"HH24:MI:SSOF') AS ts, *
FROM uniswap_v2."Pair_evt_Burn"
WHERE contract_address = '\xb4e16d0168e52d35cacd2c6185b44281ec28c9dc'
ORDER BY evt_block_number, evt_index ASC
\end{lstlisting}
}

\medskip%
\paraheader{Trades.}
{\singlespacing\small
\begin{lstlisting}[
           language=SQL,
           showspaces=false,
           basicstyle=\ttfamily,
           numbers=left,
           numberstyle=\tiny,
           commentstyle=\color{gray}
        ]
SELECT
date_trunc('minute', evt_block_time) AS minute,
SUM("amount0In") as "amount0In",
SUM("amount1In") as "amount1In",
SUM("amount0Out") as "amount0Out",
SUM("amount1Out") as "amount1Out"
FROM uniswap_v2."Pair_evt_Swap"
WHERE contract_address =  '\xb4e16d0168e52d35cacd2c6185b44281ec28c9dc'
GROUP BY 1
ORDER BY 1 ASC
\end{lstlisting}
}

\medskip%
\paraheader{Pool reserves.}
{\singlespacing\small
\begin{lstlisting}[
           language=SQL,
           showspaces=false,
           basicstyle=\ttfamily,
           numbers=left,
           numberstyle=\tiny,
           commentstyle=\color{gray}
        ]
SELECT
minute,
latest_reserves[3] AS reserve0,
latest_reserves[4] AS reserve1
FROM
(SELECT date_trunc('minute', evt_block_time) AS minute,
  (SELECT MAX(ARRAY[evt_block_number, evt_index, reserve0, reserve1]))
     AS latest_reserves
     FROM uniswap_v2."Pair_evt_Sync"
     WHERE contract_address =  '\xb4e16d0168e52d35cacd2c6185b44281ec28c9dc'
     GROUP BY 1) AS day_reserves
ORDER BY 1 ASC
\end{lstlisting}
}


\end{document}